\documentclass[10pt]{article}
\usepackage{amsmath,amsfonts,amssymb,latexsym,theorem,mathrsfs,latexsym,cite,setspace,
comment,multirow}
\usepackage{graphicx}
\usepackage{color}
\usepackage{hyperref}
\numberwithin{equation}{section}

\newcommand{\ma}[1]{\mbox{$\mathcal{#1}$}} 
\newcommand{\mas}[1]{\mbox{$\mathscr{#1}$}}

\newcommand{\D}{{\rm d}}
\newcommand{\ti}{\tilde}
\newcommand{\we}{\wedge}

\begin{document}

\begin{titlepage}


\begin{center}
{\Large \bf Kerr-Schild transformation of the Benenti-Francaviglia metric}
\end{center}
\vspace{.5cm}

\begin{center}
Masato Nozawa${}^1$ and Takashi Torii${}^2$

\vspace{.4cm}

\vskip 1cm
{\it
${}^1$Department of General Education, Faculty of Engineering, 
Osaka Institute of Technology,\\ Osaka City, Osaka 535-8585, Japan \\
${}^2$Department of System Design, Osaka Institute of Technology, Osaka City, Osaka 530-8568, Japan
}\\
{\texttt{masato.nozawa@oit.ac.jp, 
takashi.torii@oit.ac.jp}}
\end{center}

\vspace{.5cm}

\begin{abstract}
The Benenti-Francaviglia (BF) family of metrics provides the most general form of a spacetime metric that admits two mutually commuting Killing vectors and an irreducible Killing tensor. The geodesic equations for the BF family are thus completely integrable by separation of variables. Within this broad class, we explore the Kerr-Schild transformation of a degenerate subclass distinguished by the existence of a shear-free null geodesic congruence. By requiring  the deformed metric to preserve the Killing symmetry and circularity, we demonstrate that the deformed metric again falls into the degenerate BF family, modulo the replacement of a single structure function. We apply the present algorithm to ${\cal N}=2$ gauged supergravity and obtain a dyonic generalization of the Chong-Cveti\v{c}-L\"u-Pope rotating black hole solution, by taking the background metric to be a solution of the Einstein-scalar gravity. The present prescription extends to five dimensions, provided that the constant of geodesic motion associated with the extra Killing direction vanishes. The same reasoning applies to the case where the background degenerate BF metric is distorted in a (non)conformal manner. Our formalism offers a unified perspective on the relation between seed and deformed metrics in the Kerr-Schild construction. 
\end{abstract}

\vspace{.5cm}

\setcounter{footnote}{0}

\end{titlepage}

\tableofcontents

\section{Introduction}

Black holes are among the most fascinating objects predicted by general relativity, not only because of their astrophysical relevance but also due to the rich mathematical structures they embody. Among the wide variety of black hole solutions, stationary and rotating black holes occupy a particularly distinguished role,  since they would describe the final state of the dynamical process such as a gravitational collapse of a massive star. From the theoretical perspective, spinning black holes often display hidden symmetries, integrability properties, and separability structures that make them especially attractive laboratories for exploring the geometry of spacetime and the dynamics of fields in curved backgrounds.

Therefore, the construction of exact black hole solutions has been, and will continue to be, a central theme in gravitational physics. Since the Einstein equations are highly nonlinear partial differential equations, obtaining explicit solutions is a notoriously difficult task. In particular, rotating solutions involve nontrivial dependence on both radial and angular coordinates, which makes the problem even more challenging than in the static case. In the asymptotically flat and vacuum settings, remarkable progress has been achieved through solution-generating techniques, such as exploiting hidden symmetries encoded in nonlinear sigma models or employing inverse-scattering and integrability-based methods \cite{Breitenlohner:1986um,Breitenlohner:1987dg}. These approaches have enabled us to
construct  nontrivial black hole configurations in a structured fashion, as well as rotating and charged solutions in four and higher dimensions. See, for example, \cite{Iguchi:2011qi}.

However, the situation is considerably more complicated in asymptotically anti-de Sitter (AdS) spacetimes. The integrability structures that underlie the sigma-model approaches are no longer directly applicable once a cosmological constant or scalar potential are introduced~\cite{Klemm:2015uba}. As a result, it is far from obvious how to generate rotating and/or charged black hole solutions systematically in AdS backgrounds. Given the central importance of AdS spacetimes in the context of holography \cite{Maldacena:1997re}, the absence of a systematic framework for constructing exact AdS black hole solutions represents a serious limitation.

An alternative avenue for constructing exact gravitational solutions is provided by the Kerr-Schild transformation~\cite{Kerr:1965wfc,Kerr:1965vyg,Bini:2010hrs}. This technique deforms a given seed metric of Lorentz signature by adding a term quadratic in a geodesic null vector field,  multiplied by a single deformation function~\cite{Gurses:1975vu}. The method has proven remarkably powerful: both the original Kerr solution in four dimensions \cite{Kerr:1963ud} and its higher-dimensional generalizations, including the Myers-Perry(-AdS) metrics \cite{Myers:1986un, Gibbons:2004js}, can be derived within the Kerr-Schild framework. Moreover, the formalism has been successfully applied to charged solutions~\cite{Debney:1969zz,Ortaggio:2023rzp,Ayon-Beato:2025ahb} as well as to other higher-dimensional spacetimes~\cite{Dereli:1986cm,Chen:2007fs,Ortaggio:2008iq,Malek:2010mh,Srinivasan:2025hro}. For further applications, including plane wave geometries and spacetimes with perfect fluid sources, we refer the reader to~\cite{Stephani:2003tm}.

Despite these successes, the full scope of the Kerr-Schild method remains poorly understood. For instance, it is not known in general which classes of seed metrics and which types of null vectors can give rise to physically relevant solutions. In the existing literature, this issue has been investigated from the perspective of symmetry~\cite{Coll:2000rm} and through analyses based on spin coefficients~\cite{Xanthopoulos,BG1983}. A more direct and systematic relationship is clearly desirable, particularly in view of applications involving matter fields, higher dimensions with non-spherical horizon topologies, modified theories of gravity, wormholes, and related scenarios. 
Filling this gap is precisely the issue we aim to address in this work.

Inspired by above open issues, this paper focuses on the Benenti-Francaviglia (BF) metrics \cite{Benenti:1979erw,BFreview}, which provide the most general class of spacetimes admitting separability of the geodesic Hamilton-Jacobi equation. In particular, we consider the 
$D$-dimensional spacetimes with $D-2$ commuting Killing vectors. 
This family of metrics includes, as special cases, the Kerr-AdS solutions and the $D=5$ Myers-Perry-AdS black holes~\cite{Myers:1986un,Gibbons:2004js}. Within this broad class, we further restrict our attention to a distinguished subclass characterized by the existence of geodesics that keep the single non-ignorable angular coordinate fixed. In $D=4$, these geodesics correspond to shear-free null geodesic congruences and are identified with the principal null directions of the Weyl tensor. For clarity, we refer to this subclass as the degenerate BF family throughout the paper.

By exploiting the structural properties of the degenerate BF metrics, we pursue a unified description of Kerr-Schild deformations. We employ the shear-freel null geodesics of the degenerate BF metrics to put into the Kerr-Schild ansatz. We do not specify the matter fields, asymptotics and particular gravitational field equations. The only requirement imposed here is that the deformed spacetime inherits Killing symmetry and circularity. Under these conditions, we show that the deformed function can be determined uniquely and that the new metric continues to belong to the degenerate BF family. Since the only change is the replacement of the structure function $Q(q)$ by a new one, it follows that the generated solution displays the geometric behaviors inherited by the seed solution. This result represents a natural generalization of the results in~\cite{Ayon-Beato:2015nvz}, where the deformation of the Minkowski metric in terms of shear-free null geodesics was considered under the circularity condition. Our construction therefore encompasses a much broader class of cases than those analyzed in~\cite{Ayon-Beato:2015nvz}.

To illustrate the effectiveness of this framework, we apply the present procedure to charged, rotating black holes in $\mathcal{N}=2$ gauged supergravity. Our method generates a new family of solutions with additional continuous parameters, thereby extending the known solution space in a systematic fashion. The same reasoning also extends naturally to conformally related cases, where the seed metric is conformal to a degenerate BF metric, as in the Pleba\'nski-Demia\'nski family~\cite{Plebanski:1976gy}. Even in this case, the Kerr-Schild deformation still preserves the essential structural properties.

Furthermore, we generalize the construction to higher dimensions. In particular,  the degenerate BF family in $D=5$ encompasses the Myers-Perry-AdS solutions with two independent angular momenta. Our algorithm shows that these metrics can be deformed in a nonconformal way. These results collectively highlight the unifying power of the BF framework in organizing and extending the space of exact solutions.

We construct the present paper as follows. In the next section, we review the family of BF metrics and its degenerate version, which admits null geodesics with a constant angular coordinate. In section~\ref{sec:KS}, we explore the Kerr-Schild transformation of the degenerate BF metrics. Section~\ref{sec:5D} investigates the five-dimensional generalization. We conclude this paper in section~\ref{sec:conclusion}. Appendix~\ref{sec:Einstein} provides a complete classification of Einstein metrics in $D=4$ and $D=5$ within degenerate BF class. Appendix~\ref{sec:NP} presents detailed computations of the Newman-Penrose coefficients.

\section{Benenti-Francaviglia metric}
\label{sec:BF}

In \cite{BFreview}, 
Benenti and Francaviglia have obtained the necessary and sufficient conditions under which the 
$D$-dimensional metric admits the geodesic equation to be separable. 
This condition is characterized by $n(\le D)$ commuting Killing vectors and $D-n$ numbers of independent Killing tensors. In this paper, we focus on the  $D$-dimensional metric  with
$D-2$ commuting Killing vectors, for which the inverse metric is given by \cite{Benenti:1979erw}
\begin{align}
\label{Dmetric}
g^{\mu\nu}\partial_\mu \partial_\nu 
=\frac{1}{S_1(q)+S_2(p)}
\left[\left(F_1^{ij}(q)-F_2^{ij}(p)\right)
\partial_{\psi^i}\partial_{\psi^j}+Q(q)\partial_q^2 
+P(p)\partial_p^2 
\right] \,,
\end{align}
where the metric is independent of $\psi^i$ ($i=1,...,D-2$), 
and depends only on the nontrivial coordinates $q$ and $p$.
The metric is thus characterized by $(D-2)(D-1)+4$ functions 
$\{F_1^{ij}(q), S_1(q), Q(q)\}$ and $\{F_2^{ij}(p), S_2(p), P(p)\}$.

Obviously, the metric (\ref{Dmetric}) admits a set of $(D-2)$ mutually commuting Killing vector fields
\begin{align}
\label{KVs}
\xi_{(i)}=\frac{\partial}{\partial \psi^i}\,,\qquad 
\mas L_{\xi_{(i)} } g_{\mu\nu}=0 \,. 
\end{align}
These Killing vectors reflect the invariance of the metric under translations along each of the $\psi^i$ directions.

Moreover, the two-dimensional surfaces orthogonal to both $\xi_{(i)}$ are integrable. 
In other words, the distribution orthogonal to the orbits of the symmetry group generated by these commuting vector fields is surface-forming. Consequently, the metric can be brought into a block-diagonal form, where the cross terms between the Killing coordinates $\psi^i$ and the remaining nontrivial coordinates $(q, p)$ vanish. 
Spacetimes admitting such a pair of commuting Killing vector fields with orthogonal, integrable distributions are referred to as {\it circular}~\cite{Emparan:2001wk, Chrusciel:2012jk}, satisfying 
\begin{align}
\label{circularD}
\xi_{(1)}^{[\mu_1}\xi_{(2)}^{\mu_2}\cdots \xi_{(D-2)}^{\mu_{D-2}}\nabla^{\mu_{D-1}} \xi^{\mu_D]} _{(i)}=0 \,. 
\end{align}
Applying the Ricci identity and the Killing equation, 
the integrability condition of (\ref{circularD}) yields the equation for the Ricci circularity 
\begin{align}
\label{Riccicircular}
 \xi_{(i)}^\nu R_\nu{}^{[\mu } \xi_{(1)}^{\rho_1}\cdots  \xi_{(D-2)}^{\rho_{D-2}]}=0 \,. 
\end{align}
This implies that not only the metric but also the Ricci tensor is block-diagonal.
The circularity property greatly simplifies the analysis of the spacetime geometry, including the study of geodesic motion, conserved quantities, and the classification of exact solutions.

We now turn our attention to the geodesic equation $p^\nu \nabla_\nu p^\mu=0$ for this spacetime, 
where $p_\mu$ is a tangent vector of the timelike/null geodesics. 
This governs the motion of freely falling particles in this spacetime. 
To analyze geodesic motion, we employ the Hamilton-Jacobi formalism. The corresponding Hamilton-Jacobi equation takes the form
\begin{align}
\label{HJeq}
-\frac{\partial \ma S}{\partial \lambda }=\frac 12 g^{\mu\nu}\frac{\partial \ma S}{\partial x^\mu}
\frac{\partial \ma S}{\partial x^\nu}\,,
\end{align}
where $\ma S$ is  Hamilton's principal function and $\lambda$ is the affine parameter of the geodesics.  
We assume that the principal function takes the separable form 
\begin{align}
\label{HamiltonS}
\ma S=\frac 12 m^2 \lambda +\ma P_i \psi^i 
+\ma S_q(q)+\ma S_p(p) \,. 
\end{align}
Here, $m$ is the mass of the particle and 
$\ma P_i \equiv  p_\mu \xi_{(i)}^\mu$ are the 
constants of geodesic motion 
$p^\mu \nabla_\mu \ma P_i=(p^\mu \nabla_\mu p_\nu)\xi_{(i)}^\nu+p^\mu p^\nu \nabla_{(\mu}\xi_{(i)\nu)}=0$. 
Substituting this ansatz into the Hamilton-Jacobi equation (\ref{HJeq}), 
one finds that the equation is indeed separable into functions of $q$ and $p$ as
\begin{align}
\label{}
Q(q) \left(\frac{\D \ma S_q}{\D q}\right)^2+F_1^{ij}(q) \ma P_i \ma P_j  +m^2 S_1(q)=&\,\ma C \,, \label{eqq}\\
-P(p)\left(\frac{\D \ma S_p}{\D p}\right)^2+F_2^{ij}(p)\ma P_i \ma P_j -m^2 S_2(p)=&\,\ma C \,,\label{eqp}
\end{align}
where $\ma C$ is a separation constant. 
Replacing $ \nabla_\mu \ma S\to p_\mu $ 
and $m^2 \to -g^{\mu\nu}p_\mu p_\nu$, 
equation (\ref{eqp}) can be recast into
\begin{align}
\label{}
\ma C=-P(p)p_p^2 +F_2^{ij}(p)p_i p_j +S_2(p)g^{\mu\nu}p_\mu p_\nu
\equiv K^{\mu\nu} p_\mu p_\nu \,, 
\end{align}
where $K^{\mu\nu}$ is given by 
\begin{align}
\label{KT}
K^{\mu\nu}\partial_\mu \partial_\nu 
=\frac{1}{\frac1{S_1(q)}+\frac 1{S_2(p)}}
\left[\left(\frac{F_1^{ij}(q)}{S_1(q)}+\frac{F_2^{ij}(p)}{S_2(p)}\right)
\partial_{\psi^i}\partial_{\psi^j}+\frac{Q(q)}{S_1(q)}\partial_q^2 
-\frac{P(p)}{S_2(p)}\partial_p^2 
\right] \,.
\end{align}
This symmetric tensor corresponds to the Killing tensor, satisfying the defining condition 
\begin{align}
\label{}
\nabla_{(\mu }K_{\nu\rho)}=0\,.
\end{align}
This condition guarantees that there exists an additional conserved quantity along geodesics.
In particular, one can explicitly verify that the separation constant is constructed from the quadratic combination of the momentum and the Killing tensor as
\begin{align}
\label{}
p^\mu \nabla_\mu \ma C=p^\mu p^\nu p^\rho \nabla_{(\mu} K_{\nu\rho)} =0 \,.
\end{align}
Thus, the Killing tensor provides a nontrivial conserved charge, often referred to as a hidden symmetry, which is not associated with any Killing vectors.
It is worth stressing that the above derivation nowhere relies on the gravitational field equations and the spacetime dimensions; the existence of the conserved quantity is purely a consequence of the geometrical structure.
Inspecting (\ref{Dmetric}) and (\ref{KT}), one finds that the metric and the Killing tensor naturally stand on an equal footing. In particular, if the symmetric tensor (\ref{KT}) is regarded as an inverse metric, then the tensor (\ref{Dmetric}) can be interpreted as a Killing tensor \cite{Rietdijk:1995ye}. This observation highlights the deep interplay between the geometry of spacetime and its hidden symmetries.

It should be observed that the separability of the geodesic equations does not necessarily guarantee the separability of the massive Klein-Gordon equation. 
For the integrability, 
the second-order differential operator $\nabla_\mu (K^{\mu\nu}\nabla_\nu)$ 
must commute with the Klein-Gordon operator $(\nabla^2-\mu^2)$, where 
$\mu$ is the mass of the scalar field.  
 It ensures the simultaneous diagonalizability of the two operators and the existence of a common set of eigenfunctions. This leads to \cite{Carter:1977pq}
\begin{align}
\label{KTRicci}
\nabla_{\nu} \left(K^{[\mu}{}_\rho R^{\nu]\rho}\right)=0 \,. 
\end{align}
This condition  highlights that field equation separability demands more stringent geometric constraints than those required for classical particle motions.

\subsection{Degenerate BF metric}

In the following analysis for $D=4$, we mainly concentrate on
the special BF metric characterized by \cite{Demianski:1980mgt,Anabalon:2016hxg}
\begin{align}
\label{}
{\rm rank}(F_1^{ij})={\rm rank}(F_2^{ij})=1\,. 
\end{align}
In this case, the Killing coordinate part of the metric is factorized. 
We parametrize $F^{ij}_1(q)$  and $F^{ij}_2(p)$ as follows:
\begin{subequations}
\label{F12}
\begin{align}
F^{ij}_1(q)=&\,-\frac{1}{Q(q)}\left(
\begin{array}{cc}
 f_2^2(q)     & -f_1(q)f_2(q)   \\
-f_1(q)f_2(q)      &   f_1^2(q) 
\end{array}
\right)\,, \\
F^{ij}_2(p)=&\,-\frac{1}{P(p)}\left(
\begin{array}{cc}
 h_2^2(p)     & -h_1(p)h_2(p)   \\
-h_1(p)h_2(p)     &   h_1^2(p) 
\end{array}
\right)\,,
\end{align}
\end{subequations}
where we have factored out $Q(q)$ and $P(p)$ for later convenience. 
In this case, the metric reduces to  
\begin{align}
\label{metric}
\D s^2=\left[S_1(q)+S_2(p)\right] \left[
-\frac{Q(q)}{W^2}\left(h_1(p)\D \tau+h_2(p) \D \sigma \right)^2
+\frac{P(p)}{W^2}\left(f_1(q)\D \tau+f_2 (q)\D \sigma\right)^2 +\frac{\D q^2}{Q(q)}+\frac{\D p^2}{P(p)}
\right]\,, 
\end{align}
where we have denoted $\psi^i=(\tau, \sigma)$ 
and 
\begin{align}
\label{W}
W =W(p,q)\equiv h_1(p) f_2(q)-h_2(p) f_1(q)\,. 
\end{align}
We shall hereafter refer to the metric (\ref{metric}) as the {\it degenerate BF metric}.
We keep in mind that $q$ and $p$ denote the radial and angular coordinates, respectively.
Accordingly, $Q=0$ represents the loci of the Killing horizons, while $P=0$ corresponds to the axes of rotation. 
In what follows, we thus assume the following conditions
\begin{align}
\label{}
Q \ge 0 \,, \qquad P \ge 0 \,, \qquad 
S_1+S_2 >0 \,, \qquad W>0 \,. 
\end{align}

It is obvious that the metric (\ref{metric}) is invariant under the constant scale and shift transformations of $S_1$ and $S_2$ as
\begin{align}
\label{shiftS}
S_1 \to \mathsf s_0 S_1+\mathsf s_1 \,, \qquad 
S_2 \to  \mathsf s_0 S_2-\mathsf s_1 \,,
\end{align}
with 
\begin{align}
\label{}
Q\to \mathsf s_0 Q\,, \qquad 
P \to \mathsf s_0 P\,, \qquad 
\tau \to \mathsf s_0^{-1} \tau \,, \qquad 
\sigma\to \mathsf s_0^{-1} \sigma \,, 
\end{align}
where $\mathsf s_0$ and $\mathsf s_1$ are constants. 
One also finds that the metric is invariant under the ${\rm GL}(2,\mathbb R)$ symmetry 
\begin{align}
\label{GL2R}
\tau \to \mathsf a_{11}\tau+\mathsf a_{12}\sigma \,, \qquad 
\sigma \to  \mathsf a_{21}\tau+\mathsf a_{22}\sigma \,, 
\end{align}
provided that $f_{i}(q)$ and $h_{i}(p)$ are redefined as 
\begin{align}
\label{}
f_1 \to &\,  \mathsf a_{22}f_1-\mathsf a_{21}f_2 \,, \qquad 
f_2 \to  -\mathsf a_{12}f_1+\mathsf a_{11}f_2 \,, \notag \\ 
h_1 \to &\,  \mathsf a_{22}h_1-\mathsf a_{21}h_2 \,, \qquad 
h_2 \to  -\mathsf a_{12}h_1+\mathsf a_{11}h_2 \,.
\end{align}
Here, $\mathsf a_{ij}$ are constants with $\mathsf a_{11}\mathsf a_{22}-\mathsf a_{12}\mathsf a_{21}\ne 0$. 
Furthermore, 
under the rescaling 
\begin{align}
\label{}
q\to \check q(q) \,, \qquad p\to \check p(p)\,, 
\end{align}
the metric remains form-invariant up to the transformations
\begin{align}
\label{fhitr}
f_i \to \check f_i=\frac{\D \check q}{\D q} f_i \,, \qquad 
h_i \to \check h_i=\frac{\D \check p}{\D p} h_i \,, \qquad 
Q \to \check Q=\left(\frac{\D \check q}{\D q}\right)^2 Q \,, \qquad 
P \to \check P=\left( \frac{\D \check p}{\D p}\right)^2 P \,. 
\end{align}
Each of these transformations will be used for the classification of Einstein spaces.

For the  degenerate BF metric (\ref{metric}), the computation of the Ricci tensor becomes considerably more tractable.
To proceed, we introduce the following orthonormal frame
\begin{align}
\label{}
e^0=&\, \frac{\sqrt{(S_1+S_2)Q}}{W}(h_1 \D \tau +h_2 \D \sigma)\,, 
&
e^1=&\, \frac{\sqrt{(S_1+S_2)P}}{W}(f_1 \D \tau +f_2 \D \sigma)\,, 
\notag \\
e^2=&\, \frac{\sqrt{S_1+S_2}}{\sqrt{Q}}\D q \,, 
&
e^3=&\, \frac{\sqrt{S_1+S_2}}{\sqrt{P}} \D p \,, 
\end{align}
with the local Lorentz metric $\eta_{ab}={\rm diag}(-1,1,1,1)$. 
In this local frame, the Ricci circularity condition (\ref{Riccicircular}) implies 
\begin{align}
\label{}
R_{02}=R_{03} =R_{12}=R_{13}=0 \,. 
\end{align}
Furthermore,  it is also useful for later use to record
\begin{align}
\label{Rpq}
R_{23}=&\, \frac{3\sqrt{QP}}{2(S_1+S_2)}\partial_p \partial_q \left[\ln \left(\frac{W}{S_1+S_2}\right)\right]\,, \\
R^0{}_0+R^1{}_1=&\, -\frac{W}{(S_1+S_2)^2}
\left[\partial_p \left(\sqrt{\frac{P}{Q}}\partial _p \rho \right)+
\partial_q \left(\sqrt{\frac{Q}{P}}\partial _q \rho \right)\right]\,,
\end{align}
where 
\begin{align}
\label{}
\rho\equiv \frac{S_1+S_2}{W}\sqrt{P(p)Q(q)}\,.   
\end{align}

By exploiting the freedom (\ref{fhitr}), we can always set  e.g., $f_1(q)=h_1(p)=1$. 
In this gauge, the frame component $R_{01}$ 
of the Ricci tensor is simplified to
\begin{align}
\label{R01g}
R_{01}= \frac{ \sqrt{QP}}{2(S_1+S_2)^2} \left[\partial_q \left(\frac{S_1+S_2}{f_2-h_2}f_2'(q)\right)
+\partial_p \left(\frac{S_1+S_2}{f_2-h_2}h_2'(p)\right)\right]\,.
\end{align}
Note that the Petrov type of the degenerate BF metric (\ref{metric}) is generically I \cite{Galtsov:2024vqo}.

\subsection{Examples}

Since the degenerate BF metric (\ref{metric}) still contains as many as eight arbitrary functions, it is far from easy to extract its physical properties in a straightforward way. To gain a clearer understanding, it is therefore instructive to restrict ourselves to certain representative subclasses. In the following, we shall examine these specific cases in detail and discuss the physical properties that emerge from them.

\subsubsection{Carter family}

Setting
\begin{align}
\label{}
f_1(q)=h_1(p)=1\,, \qquad 
S_1(q)=f_2(q)=q^2 \,, \qquad 
S_2(p)=-h_2(p)=p^2 \,, 
\end{align}
we obtain the Carter family of metrics \cite{Carter:1973rla}
\begin{align}
\label{Carter}
\D s^2=-\frac{Q(q)}{q^2+p^2}(\D \tau-p^2 \D \sigma)^2
+\frac{P(p)}{q^2+p^2}(\D \tau+q^2 \D \sigma)^2
+(q^2+p^2)\left(\frac{\D q^2}{Q(q)}+\frac{\D p^2}{P(p)}\right)\,.
\end{align}
The Carter family of metrics is specified by two arbitrary functions $Q$ and $P$, and is
of Petrov D. 
This includes the Kerr-(A)dS metric, which satisfies the vacuum Einstein's equations
with a  cosmological constant $R_{\mu\nu}=\Lambda g_{\mu\nu}$. 
A review of the classification of Einstein spaces, together with the transformation to the more familiar Boyer-Lindquist coordinates, is provided in Appendix~\ref{sec:Einstein}.

It is worth emphasizing that the Killing tensor associated with the Carter family of metrics can be constructed out of a  Killing-Yano tensor $f_{\mu\nu}=f_{[\mu\nu]}$~\cite{Yano} (see e.g., \cite{Yasui:2011pr} for review).
Specifically, one may write (\ref{KT}) as 
\begin{align}
\label{KTKY}
 K_{\mu\nu}= - f_{\mu \rho}f_\nu{}^\rho \,, 
\end{align}
where $f_{\mu\nu}$ satisfies the Killing-Yano equation
\begin{align}
\label{}
\nabla_{(\mu} f_{\nu)\rho }=0\,. 
\end{align}
An explicit expression for the two-form is given by
\begin{align}
\label{}
\frac 12 f_{\mu\nu}\D x^\mu \we \D x^\nu =p(\D \tau-p^2 \D \sigma) \we \D q -q \D p \we (\D \tau +q^2 \D \sigma) \,.  
\end{align}
This construction makes it manifest that the Killing-Yano tensor is in fact the ``square-root'' of the Killing  tensor.

When the Killing tensor is expressed in the form (\ref{KTKY}), 
one can check that the Killing tensor and the Ricci tensor commute
as (1,1) index tensors $K^{[\mu}{}_\rho R^{\nu]\rho}=0$. 
As a consequence, the compatibility condition (\ref{KTRicci}) is automatically satisfied. Therefore, the separability of the massive Klein-Gordon equation is guaranteed for the Carter family. Furthermore, test Dirac field equation on the Carter background is also separable, thereby providing a direct link between the hidden symmetry generated by the Killing-Yano tensor and the integrability of matter field equations~\cite{Yasui:2011pr,Carter:1979fe}. 

When the vacuum Einstein's equations are satisfied, the 
Carter solution is alternatively characterized by the vanishing Simon tensor~\cite{Simon,Mars:1999yn,Nozawa:2021udg}.
The Simon tensor is constructed from the sigma-model variables associated with the timelike Killing field, which encode the alignment between the spacetime curvature and the underlying symmetries. Its vanishing condition provides an invariant characterization that singles out the Kerr (or more generally, Carter) geometry within the class of stationary vacuum spacetimes.

\subsubsection{Chong-Cveti\v{c}-L\"u-Pope solution}

The next example is the nonextremal black hole solution to the ${\cal N}=2$ supergravity with 
the prepotential $F(X)=-i X^0X^1$. The Lagrangian of this theory is given by 
\begin{align}
\ma L=&\, \frac 12(R-2V ) \star 1-\frac{\D z \we \star \D \bar z}{(z+\bar z)^2}
-\frac 14 (z+\bar z)F^0 \we\star F^0-\frac{z+\bar z}{4z\bar z}F^1 \we \star F^1
+\frac{z-\bar z}{4i} F^0 \we F^0-\frac{z-\bar z}{4i z\bar z}F^1\we F^1\,,
\label{N2Lag1}
\end{align}
where $F^I=\D A^I$ ($I=0, 1$) are the electromagnetic gauge fields, and  
$z=e^{-\phi}+i\chi $ is a complex scalar field parametrizing the ${\rm SU}(1,1)/{\rm U}(1)$ coset space. 
The potential is given by\footnote{Here we focus only on the case in which the theory admits an AdS vacuum,
which occurs when the two gauge coupling constants have the same sign. 
Within this setup, the electromagnetic field can be rescaled so that only a single coupling constant $g$ remains.
See, e.g., \cite{Nozawa:2022upa}.}
\begin{align}
\label{}
V(z, \bar z)=-g^2\left(2+\frac{1+|z|^2}{z+\bar z}\right)
=-g^2\left(2+\cosh \phi+\frac 12 e^\phi \chi^2 \right)\,.
\end{align}
The origin $\phi=\chi=0$ is an AdS vacuum
$\partial_\phi V=\partial_\chi V=0$ with $V_{\phi=\chi=0}=-3g^2$,
where $g$ represents the inverse AdS radius. 

Einstein's equations derived from (\ref{N2Lag1}) are written as 
\begin{align}
\label{N2Eineq}
E_{\mu\nu} \equiv R_{\mu\nu}-T^{(z)}_{\mu\nu}-T^{(F)}_{\mu\nu} =0 \,,  
\end{align}
where (trace-reversed) stress-energy tensors are given by
\begin{align}
\label{}
T^{(z)}_{\mu\nu}=&\,\frac{2}{(z+\bar z)^2}\nabla_{(\mu }z\nabla_{\nu)}\bar z +V g_{\mu\nu} \,, \\
T^{(F)}_{\mu\nu}=&\, \frac {z+\bar z}{2} \left((F^0)_{\mu\rho}(F^0)_{\nu}{}^\rho -\frac 14 g_{\mu\nu}(F^0)^2 \right)
+\frac{z+\bar z}{2z\bar z}\left((F^1)_{\mu\rho}(F^1)_{\nu}{}^\rho -\frac 14 g_{\mu\nu}(F^1)^2 \right)\,.
\end{align}
The equations for the gauge fields are
\begin{align}
\label{N2Maxeq}
\D F^I=0 \,, \qquad \D H_I=0\,,   
\end{align}
 where 
\begin{align}
\label{}
H_0 \equiv -\frac{z+\bar z}{2}\star F^0+\frac{z-\bar z}{2i}F^0 \,, \qquad 
H_1 \equiv -\frac{z+\bar z}{2z\bar z}\star F^1-\frac{z-\bar z}{2i z\bar z}F^1 \,. 
\end{align}
The complex scalar field $z$ obeys
\begin{align}
\label{N2zeq}
\nabla^2 z-\frac{2}{z+\bar z}(\nabla z)^2-(z+\bar z)^2 \partial_{\bar z} V
-\frac{(z+\bar z)^2}{8}\left(F^0_{\mu\nu}(F^0+i \star F^0)^{\mu\nu} 
-\frac{1}{\bar z^2}F^1_{\mu\nu}(F^1+i \star F^1)^{\mu\nu} 
\right)=0\,.
\end{align}

A stationary, rotating, non-extremal AdS black hole solution of this theory with nontrivial scalar hair was first obtained by Chong-Cveti\v{c}-L\"u-Pope (CCLP)~\cite{Chong:2004na} and later generalized by~\cite{Gnecchi:2013mja}. Here, we focus on the magnetically charged case discussed in~\cite{Gnecchi:2013mja}, which can be recognized as belonging to the degenerate BF class~(\ref{metric}) by choosing
\begin{align}
\label{}
f_1(q)=h_1(p)=1\,, \qquad S_1(q)=f_2(q)=q^2-q_0^2 \,,  \qquad S_2(p)=-h_2(p)=p^2 \,, 
\end{align}
where $q_0$ is a constant. With this parametrization, the CCLP solution takes the explicit form
\begin{align}
\label{CCLP}
\D s^2 = -\frac{Q(q)}{q^2-q_0^2+p^2}(\D \tau-p^2 \D \sigma)^2+\frac{P(p)}{q^2-q_0^2+p^2}(\D \tau+(q^2-q_0^2) \D \sigma)^2+(q^2-q_0^2+p^2) \left(
\frac{\D q^2}{Q(q)}+\frac{\D p^2}{P(p)}\right)\,,
\end{align}
with the structure functions
\begin{align}
\label{}
Q(q)=&\, \mathtt c_2+\frac 12 (\mathtt P_0^2+\mathtt P_1^2)
+(q^2-q_0^2)[\mathtt c_1+g^2(q^2-q_0^2)]+\frac{\mathtt P_0^2-\mathtt P_1^2}{2q_0}q \,, 
\\
P(p)=&\, \mathtt c_2 +p^2 (-\mathtt c_1+g^2 p^2) \,.
\end{align}
Here, $\mathtt c_1$, $\mathtt c_2$, $\mathsf P_1$, $\mathsf P_2$ are constants. 
The corresponding gauge fields and scalar fields are given by
\begin{align}
\label{}
A^I= \frac{\mathtt P_Ip}{q^2-q_0^2+p^2} \left(\D \tau+(q^2-q_0^2)\D \sigma\right)\,, \qquad 
z =\frac{q+q_0+i p}{q-q_0+i p}\,.
\end{align}
To make contact with the parametrization employed in \cite{Chong:2004na}, one must perform an electromagnetic duality rotation on $F^1$, and introduce the redefinitions $q=r +m(s_1^2+s_2^2)$, 
$p=a\cos\theta$, $\tau=t-a \phi$, $\sigma=-\phi/a$ with $q_0=m(s_1^2-s_2^2)$, $\mathtt c_1=1+a^2g^2$, 
$\mathtt c_2=a^2$ and
$s_i =\sinh \delta_i $. 

Although this solution~(\ref{CCLP}) lies outside the Carter family and belongs to Petrov type I, it nonetheless admits nontrivial separability properties. In particular, both of the geodesic and scalar field equations were shown to be separable in \cite{Vasudevan:2005bz,Houri:2010fr}.

\subsection{Shear-free null geodesic congruence}

For the degenerate BF metric, the separated Hamilton-Jacobi equations (\ref{eqq}) and (\ref{eqp}) lead to 
\begin{align}
\label{}
Q(q)\left(\frac{\D \ma S_q}{\D q}\right)^2
-\frac{(f_2 (q)\ma P_\tau-f_1(q)\ma P_\sigma)^2}{Q(q)}+m^2 S_1(q)=&\,  \ma C \,, \label{sep1}\\
P(p)\left(\frac{\D\ma S_p}{\D p}\right)^2
+\frac{(h_2 (p)\ma P_\tau-h_1(p)\ma P_\sigma)^2}{P(p)}+m^2 S_2(p)=&\, -\ma C \,.
\label{sep2}
\end{align}
Here, we focus on the special class of null geodesics ($m=0$) with
\begin{align}
\label{PNC0}
\ma C=0 \,.
\end{align}
Inspecting (\ref{sep2}) and noting that  $P(p)\ge 0$, 
the following conditions must be satisfied simultaneously
\begin{align}
\label{PNC2}
h_2 (p)\ma P_\tau -h_1(p)\ma P_\sigma=0 \,, \qquad \frac{\D \ma S_p}{\D p}=0  \,.
\end{align}
It follows that the angular coordinate $p$ ($\sim \cos\theta$) remains constant
along this family of null geodesics. Solving (\ref{sep1}), 
the tangent vectors of the null geodesics are now given by 
\begin{align}
\label{null}
k_\mu^{(\pm)} \D x^\mu =-h_1(p)\D \tau-h_2(p) \D \sigma \pm \frac{W}{Q(q)}\D q \,. 
\end{align}
It can be easily checked that both families of null geodesics are {\it shear-free} \cite{Anabalon:2016hxg,Galtsov:2025nia}
\begin{align}
\label{shearfree}
 \sigma^{(\pm)}_{\mu\nu}\equiv K^{(\pm)}_{\mu\nu}-\frac 12 K^{(\pm)} h _{\mu\nu}=0\,.
\end{align}
Here, we have defined $K^{(\pm)}_{\mu\nu}=h_\mu{}^\rho h_\nu {}^\sigma \nabla_{(\rho}k^{(\pm)}_{\sigma)} $, 
$K^{(\pm)}=g^{\mu\nu} K^{(\pm)}_{\mu\nu}$ and 
$h_\mu{}^\nu=\delta_\mu{}^\nu+k^{(\pm)}_\mu n^{(\mp)\nu} +n^{(\mp)}_\mu k^{(\pm)\nu}$, where 
$n^{(\pm)}_\mu =Q(q)[S_1(q)+S_2(p)]k_\mu^{(\mp)}/(2W^2)$.
We will see in the next section that these geodesics play a significant role in the Kerr-Schild deformation.

One can also verify that 
the null geodesics (\ref{null}) correspond to the repeated principal null directions of the Weyl tensor, satisfying
\begin{align}
\label{Weyleig}
k_{[\alpha}^{(\pm)} C_{\mu]\nu\rho [\sigma} k_{\beta]}^{(\pm)} k^{(\pm) \nu} k^{(\pm) \rho}=0 \,. 
\end{align}
In the notation of Newman-Penrose formalism, this means that the Weyl scalars $\Psi_0$ and $\Psi_4$ vanish $\Psi_0=\Psi_4=0$.  
It must be stressed that  the simultaneous fulfillment of conditions (\ref{PNC0}), (\ref{shearfree}) and (\ref{Weyleig}) occurs only in $D=4$, as we will see in section~\ref{sec:5D}.

\section{Kerr-Schild transformation}
\label{sec:KS}

We now proceed to the central part of our analysis, where we examine how the Kerr-Schild transformation operates on the degenerate BF metric (\ref{metric}) and discuss the resulting implications.

\subsection{Formalism}

Before going into the detail of our main results, 
we provide a concise review of the Kerr-Schild formalism \cite{Kerr:1965wfc,Kerr:1965vyg,Bini:2010hrs}
(cf, section 32 of \cite{Stephani:2003tm}). 
Let $k_\mu$ be a tangent vector of null geodesics for the $D$-dimensional background metric $g_{\mu\nu}$ as
\begin{align}
\label{}
g^{\mu\nu} k_\mu k_\nu =0 \,, \qquad k^\nu \nabla_\nu k^\mu =0 \,.   
\end{align}
In terms of $k_\mu$, 
let us consider the following deformation of the metric
\begin{align}
\label{KerrSchild}
\ti g_{\mu\nu} =g_{\mu\nu}+2 H k_\mu k_\nu \,, 
\end{align}
where $H$ is a scalar function to be determined. 
In the context of modified gravity, such a metric $\ti g_{\mu\nu}$ is termed the ``disformal metric,'' precisely because the relation between the two metrics cannot be reduced to a mere conformal transformation.
On account of 
\begin{align}
\label{}
\ti g^{\mu\nu}= g^{\mu\nu}-2H k^\mu k^\nu \,, \qquad \sqrt{-\ti g}=\sqrt{-g}\,, 
\end{align}
one can verify that 
 $k_\mu$ is also a tangent vector of null geodesics with respect to the deformed metric 
 \begin{align}
\label{}
\ti g^{\mu\nu} k_\mu k_\nu=0 \,, \qquad 
k^\nu \ti \nabla_\nu k^\mu =0 \,.
\end{align}
Here, 
$\ti \nabla_\mu$ denotes the covariant derivative with respect to $\ti g_{\mu\nu}$, and 
 we do not distinguish $\ti k^\mu=\ti g^{\mu \nu}k_\nu=g^{\mu \nu}k_\nu=k^\mu$ for notational simplicity. 
Writing the deviation from the background metric as $h_{\mu\nu}=2H k_\mu k_\nu$, 
one can show that the Ricci tensor $\ti R^\mu{}_\nu$ of $\ti g_{\mu\nu}$
is related to the one $R^\mu{}_\nu$ of $g_{\mu\nu}$ as 
\begin{align}
\label{RicciKS}
\ti R^\mu{}_\nu =R^\mu{}_\nu+{}^{(1)\!}R^\mu{}_\nu\,, 
\end{align}
where ${}^{(1)\!}R^\mu{}_\nu$ denotes the linearized Ricci tensor for $h_{\mu\nu}$ 
\begin{align}
\label{}
{}^{(1)\!}R^\mu{}_\nu\equiv 
\frac 12 \left(-\nabla^\rho\nabla_\rho h^\mu{}_\nu
+\nabla_\rho \nabla^\mu h^\rho{}_\nu +\nabla_\rho
\nabla_\nu h^{\rho \mu} \right)-h^{\mu}{}_\rho R^\rho{}_{\nu} \,. 
\end{align}
What is remarkable here is that the Ricci tensor becomes automatically linearized, despite the fact that we have nowhere assumed $h_{\mu\nu}$ to be infinitesimally small \cite{Dereli:1986cm}.
It is worth noting that the linearization holds only when the Ricci tensor is written in the mixed index form, with one index raised and the other lowered.
Therefore,  $\ti g_{\mu\nu}$ satisfies the vacuum Einstein's equations
with a cosmological constant $\ti R_{\mu\nu}=\frac{2}{D-2}\Lambda \ti g_{\mu\nu}$, 
provided that  $h_{\mu\nu}$ satisfies the linearized Einstein's equations ${}^{(1)\!}R^\mu{}_\nu=0$ with respect to the background metric $g_{\mu\nu}$. 

This is the procedure how the explicit Kerr(-AdS) metric $\ti g_{\mu\nu}$ is constructed out of the AdS metric $g_{\mu\nu}$~\cite{Kerr:1963ud,Carter:1973rla}, which we shall now overview. 
The background $D=4 $ AdS metric can be written as
\begin{align}
\label{}
\D s^2=&\,-\frac{(1+g^2 r^2)(1-a^2g^2\cos^2\theta)}{\Xi}\D \bar t^2+\frac{(r^2+a^2)\sin^2\theta}{\Xi}\D \bar\phi^2
\notag \\
&\,+(r^2+a^2\cos^2\theta)\left(\frac{\D r^2}{(r^2+a^2)(1+g^2r^2)}+\frac{\D \theta^2}{1-a^2g^2\cos^2\theta}\right)\,,
\label{AdS0}
\end{align}
where $a$ is a constant and $\Xi=1-a^2g^2$.
The metric satisfies $R_{\mu\nu}=-3g^2 g_{\mu\nu}$, where 
$g$ represents the reciprocal of AdS radius. In the $g\to 0$ limit, the Minkowski spacetime is recovered.
The above form (\ref{AdS0}) of the metric explicitly exhibits the Cartan subgroup ${\rm SO}(2)\times {\rm SO}(2)$ of  the full isometry group ${\rm SO}(3,2)$.
To illustrate this, it is convenient to consider the hypersurface 
$-(X^0)^2-(X^1)^2+(X^2)^2+(X^3)^2+(X^4)^2=-g^{-2}$ in a five dimensional space 
$\D s^2=-(\D X^0)^2-(\D X^1)^2+(\D X^2)^2+(\D X^3)^2+(\D X^4)^2$. 
The embedding is given by
\begin{subequations}
\begin{align}
\label{}
X^0+i X^1=&\,\frac1g\sqrt{\frac{(1+g^2r^2)(1-a^2g^2\cos^2\theta)}{1-a^2g^2}} e^{ig \bar t}\,,   \\ 
X^2+i X^3=&\,\sqrt{\frac{r^2+a^2}{1-a^2g^2}}\sin\theta e^{i\bar\phi} \,, \\
X^4=&\, r \cos\theta \,. 
\end{align}
\end{subequations}
It follows that the prefactor $\Xi$ appearing in $g_{\bar t\bar t}$ and $g_{\bar \phi\bar \phi}$ ensures that both  $g\bar t$ and $\bar\phi$ have $2\pi $ period, corresponding to the rotations in $(X^0, X^1)$ and $(X^2,X^3)$ planes.\footnote{To eliminate the closed timelike curves generated by $\partial/\partial \bar t$, one typically considers the universal covering space of AdS, which is what we usually mean when referring to ``AdS.''} 

The tangent vectors of the shear-free null geodesics are given by
\begin{align}
\label{}
k_\mu^{(\pm)} \D x^\mu=-\frac{1-a^2g^2\cos^2\theta}{\Xi}\D \bar t\pm \frac{r^2+a^2\cos^2\theta}{(r^2+a^2)(1+g^2r^2)}\D r
+\frac{a\sin^2\theta}{\Xi}\D \bar \phi \,. 
\end{align}
For the sake of simplicity, we consider $k_\mu =k^{(+)}_\mu$ as a deformation vector. 
Solving the linearized Einstein's equations ${}^{(1)\!} R^\theta{}_\theta={}^{(1)\!} R^\theta{}_{\bar t}={}^{(1)\!} R^\theta{}_{\bar \phi}=0$ together with their integrability conditions, 
the general solution $H=H(\bar t, \bar \phi , r, \theta)$ is found to be
\begin{align}
\label{}
H=H(r,\theta)=\frac{Mr}{r^2+a^2\cos^2\theta} \,, 
\end{align}
where $M$ is an integration constant. 
By a further coordinate transformation
\begin{align}
\label{tphiKerrAdS}
\D \bar t=\D t-\frac{2Mr}{(1+g^2r^2)\Delta_r(r)}\D r\,, \qquad 
\D \bar \phi=\D \phi-\frac{2a Mr}{(r^2+a^2)\Delta_r(r)}\D r+a g^2 \D t \,, 
\end{align}
one derives the Kerr-AdS metric in the Boyer-Lindquist coordinates~\cite{Carter:1973rla}
\begin{align}
\label{}
\D s^2=&\,-\frac{\Delta_r(r)}{r^2+a^2\cos^2\theta}\left(\D t-\frac{a\sin^2\theta}{\Xi}\D \phi \right)
^2+\frac{\Delta_\theta(\theta)\sin^2\theta}{r^2+a^2\cos^2\theta}\left(\frac{r^2+a^2}{\Xi}\D \phi
-a \D t\right)^2
\notag \\
&+(r^2+a^2\cos^2\theta)\left(\frac{\D r^2}{\Delta_r(r)}+\frac{\D \theta^2}{\Delta_\theta(\theta)}\right)\,, 
\label{KerrAdS}
\end{align}
where 
$\Delta_r(r)\equiv (1+g^2r^2)(r^2+a^2)-2Mr$ and $\Delta_\theta(\theta)\equiv 1-a^2g^2\cos^2\theta$. 

Although these computations themselves pose no difficulty, 
the Kerr-Schild transformation starting from the metric (\ref{AdS0}) raises several fundamental questions from both physical and geometrical perspectives.
To begin with, it is far from obvious why the coordinate system employed in (\ref{AdS0})--among the many possible coordinate representations of AdS--should be regarded as the natural or preferred choice for deriving the Kerr-AdS metric (\ref{KerrAdS}). Understanding the rationale for this choice is essential, since the structure of the Kerr-Schild ansatz is highly sensitive to the coordinate system adopted for the seed metric.

The second concern arises in connection with the coordinate transformation (\ref{tphiKerrAdS}) to the Boyer-Lindquist form. While this step is often taken to recast the Kerr-AdS solution into a more familiar representation, it has the drawback of obscuring the geometric relation between the shear-free null geodesics in the original and transformed coordinates. Although the transformation rules for the spin coefficients and curvature components, given in Appendix~\ref{sec:NP}, are well established, the direct connection between the Kerr-Schild congruence and the separability structure of the geodesic equations becomes less transparent after the transformation.

Finally, when the seed metric is chosen to be the Schwarzschild-AdS solution with $r$-$\theta$ separable shear-free null geodesics, the Kerr-Schild transformation does not yield the expected rotating Kerr-AdS family. Instead, it generates a new family of Schwarzschild-AdS metrics with different values of the mass parameter. This outcome raises the question of whether the construction genuinely captures the rotational generalization or merely reshuffles the parameters of the static solution. Addressing these difficulties is the primary objective of the next section.

\subsection{Kerr-Schild transformation of the degenerate BF metric}

To resolve the issues raised above, we consider the Kerr-Schild transformation
of the degenerate BF metric (\ref{metric})  
by employing the shear-free null geodesic vector (\ref{null})
 as a deformation null vector.  This choice is far from arbitrary. The degenerate BF metric offers a simple but nontrivial background in which the geodesic equation can be solved analytically. Such solvability of the geodesic motion constitutes a necessary prerequisite for a consistent implementation of the Kerr-Schild transformation. Moreover, the use of a coordinate representation that makes separability explicit reveals the crucial role played by the shear-free null geodesics.

In performing the Kerr-Schild transformation, 
we only consider $k_\mu=k_\mu^{(+)}$ for simplicity, and 
require that the vector fields defined in (\ref{KVs}) remain Killing vectors of the deformed metric $\tilde{g}_{\mu\nu}$, namely the deformation function $H$ should be $H = H(p, q)$. 
In addition to preserving the Killing symmetry, we also demand that the deformed metric $\tilde{g}_{\mu\nu}$  satisfies the circularity condition (\ref{circularD}). In $D=4$, the circularity condition boils down to  
\begin{align}
\label{circular}
\xi _{(i)}^\mu \omega_{(j)\mu}=0\,, \quad (i\ne j )
\end{align}
where $\omega_{(i)\mu}$ are twist vectors defined by 
\begin{align}
\label{}
\omega_{(i)\mu}\equiv \epsilon_{\mu\nu\rho\sigma} \xi^\nu_{(i)}\nabla^\rho \xi^\sigma_{(i)} \,. 
\end{align}
In view of our ultimate goal of obtaining a rotating black hole solution, these two requirements are  physically well motivated. 
The circularity-preserving deformation of the Minkowski seed was first analyzed in \cite{Ayon-Beato:2015nvz}. In the present paper, we extend their argument to a broader framework by utilizing the degenerate BF metrics.

Notably, the only possible form of $H = H(p, q)$ compatible with above two conditions is
uniquely fixed to be 
\begin{align}
\label{solH}
H(p, q) = \frac{[Q(q)-\ti Q(q)][S_1(q)+S_2(p)]}{2 W(p,q)^2 } \,,
\end{align}
where $\ti Q(q)$ is an arbitrary function of $q$. 
With this deformation, there appear the cross terms $\D q \D \tau$ and $\D q \D \sigma$ in the metric,
which can be made to vanish by the following coordinate transformations 
\begin{align}
\label{newtausigma}
\ti \tau =\tau +\int f_2(q) \left(\frac 1{\ti Q(q)}-\frac 1{Q(q)}\right)\D q \,, \qquad 
\ti \sigma =\sigma -\int f_1(q) \left(\frac 1{\ti Q(q)}-\frac 1{Q(q)}\right)\D q\,.
\end{align}
In terms of these new coordinates, it turns out that 
the deformed metric $\D \ti s^2=\ti g_{\mu\nu}\D \ti x^\mu \D \ti x^\nu$ 
($\ti x^\mu =(\ti \tau, \ti \sigma, q, p)$) takes the standard circular form, and, fairly remarkably, is found to fall once again within the class of degenerate BF metrics
\begin{align}
\label{metric0}
\D \ti s^2=[S_1(q)+S_2(p)] \left[
-\frac{\ti Q(q)}{W^2}\left(h_1(p)\D\ti \tau+h_2(p) \D\ti\sigma \right)^2
+\frac{P(p)}{W^2}\left(f_1(q)\D \ti \tau+f_2 (q)\D \ti \sigma\right)^2 +\frac{\D q^2}{\ti Q(q)}+\frac{\D p^2}{P(p)}
\right]\,.
\end{align}
This amounts to replacing the structure function $Q(q)$ with $\ti Q(q)$. 
This replacement is also manifest in the expression of the principal null vector 
$\ti k_\mu =k_\nu (\partial x^\nu/\partial \ti x^\mu)$, which  
retains the same functional form as $k_\mu$ given by (\ref{null}): 
\begin{align}
\label{}
\ti k_\mu \D \ti x^\mu =-h_1(p)\D\ti \tau-h_2(p) \D \ti\sigma +\frac{W}{\ti Q(q)}\D q \,. 
\end{align}
Therefore, it is obvious that $\ti k_\mu$ is a tangent vector to the shear-free null geodesics with respect to $\ti g_{\mu\nu}$, 
\begin{align}
\label{}
\ti g^{\mu\nu} \ti k_\mu \ti k_\nu  =0 \,, \qquad 
\ti k^\nu \ti\nabla_\nu \ti k^\mu=0 \,, \qquad 
\ti \sigma_{\mu\nu}=0\,. 
\end{align}
Analogously, the Killing tensor $\ti K^{\mu\nu}$ for the metric $\ti g_{\mu\nu}$ is obtained  just by $Q(q)\to \ti Q(q)$
in (\ref{KT}) and (\ref{F12}).  
It follows that the background and the deformed metrics do not display any essential geometric distinction.

The considerations outlined above allow us to address, in a systematic manner, the questions raised in the previous subsection.
In order to obtain a stationary black hole metric, it is legitimate to impose the requirement that the resulting spacetime be circular. Indeed, under fairly physically reasonable assumptions, any stationary black hole must admit an axisymmetric Killing vector~\cite{Hawking:1971vc,Hollands:2006rj} and thus belong to the circular class. Circularity is therefore not merely a technical convenience but a fundamental structural property of realistic stationary black hole geometries.

This framework makes it clear that the seed metric must lie within the degenerate BF class in order for the deformed metric to admit geodesic separability. Moreover, the present construction clarifies the role of the coordinate transformation (\ref{newtausigma}), which maps the degenerate BF metric into itself. This stands in sharp contrast to the situation in Boyer-Lindquist coordinates, where the significance of the transformation (\ref{tphiKerrAdS}) remains rather obscure. The present analysis thus provides a consistent and direct geometric explanation of how both the shear-free congruence and the separability structure are preserved under the transformation.

Finally, it turns out that  the seed and generated metrics must share the same cross-term structure, namely of the form $h_1(p)\D \tau + h_2(p)\D \sigma$ and $f_1(q)\D \tau + f_2(q)\D \sigma$.
This observation explains why neither the global patch nor the Poincar\'e patch of AdS leads to rotating solutions: when expressed in the (conformally) degenerate BF form, these representations do not admit off-diagonal terms. In contrast, the metric (\ref{AdS0}) can be recast into the Carter form (\ref{Carter}) simply by performing a linear mixing of the two Killing coordinates
$\tau=(\bar t+a \bar\phi)/\Xi$, $\sigma=(\bar \phi+a g^2 \bar t)/(a\Xi)$, thereby admitting off-diagonal terms. The same reasoning clarifies why the Schwarzschild-AdS seed--characterized by the absence of such cross terms $f_1 = h_2 = 0$--fails to generate the Kerr-AdS family. Instead of producing rotation, the transformation merely modifies the mass parameter of the static solution. By contrast, the Minkowski/AdS metric can be recast into the Carter form (\ref{Carter}) with nonvanishing off-diagonal terms. Although these terms for Minkowski/AdS arise solely from adopting a rotating coordinate frame, and hence represent only fictitious rotations, they nevertheless prove to be essential in the present formalism.

\subsubsection{Application to gauged supergravity}

As a concrete example of the above formulation, we demonstrate the derivation and the generalization of the 
CCLP solution (\ref{CCLP}) via the Kerr-Schild formalism.

As discussed in \cite{Ett:2010by, Malek:2014dta}, the CCLP solution (\ref{CCLP}) does not fit into a Kerr-Schild ansatz if the background is taken to be pure AdS. As clarified in appendix~\ref{sec:Einstein}, we show that any Einstein metric within the degenerate BF family necessarily belongs to the Carter class. Since the CCLP solution lies outside the Carter class, it cannot be generated from a pure AdS seed by a Kerr-Schild deformation. This is because the Kerr-Schild procedure preserves the cross-term structures of the form $h_1 \D \tau + h_2 \D \sigma$ and $f_1 \D \tau + f_2 \D \sigma$, and the specific functional forms of $f_2$ and $h_2$ that arise in the Carter subclass are incompatible with those in the CCLP geometry. Consequently, a pure AdS background is not a viable seed for deriving the CCLP solution. Of course, once the full solution is known, one can retrospectively infer what background it corresponds to. However, the true utility of the Kerr-Schild construction lies in generating black hole solutions directly from a prescribed background, and for this purpose it is desirable that a suitable background metric be available from the outset.

In a recent work \cite{Hassaine:2024mfs}, it was argued that the CCLP solution can be embedded into the Kerr-Schild ansatz, provided the background is taken to be a corresponding extremal black hole. However, this prescription also comes at a price, since one must solve the complete set of field equations (\ref{N2Eineq}), (\ref{N2Maxeq}), (\ref{N2zeq}) in order to determine the background extremal geometry itself. Even if the extremal configuration is assumed to preserve supersymmetry, this remains a formidable task, as the BPS conditions in gauged supergravity still lead to a highly nontrivial system of nonlinear equations~\cite{Cacciatori:2008ek,Klemm:2011xw}. Thus, also in this case, the background can only be identified a posteriori, once the full solution is already known, unless the extremal solutions are obtained in full generality. This stands in contrast to the standard Kerr-Schild approach, where the background metric is specified a priori and black hole solutions are generated from it. 

In light of these difficulties, we propose instead that the background metric for generating the CCLP solution (\ref{CCLP}) should be taken not as pure AdS, but as a solution to Einstein-scalar gravity satisfying $R^\mu{}_\nu = T^{(z)\mu}{}_\nu$ with $F^{(I)}=0$. Such a self-gravitating scalar background provides a more natural and consistent framework, since the scalar fields are dynamically evolving and thus furnish the appropriate off-diagonal structure related to the scalar charge, which is 
required to support the Kerr-Schild deformation leading to the CCLP solution.

To obtain the background seed solution, we restrict ourselves to the subclass characterized by
$S_1=f_2$, $S_2=-h_2$, together with the gauge choice $f_1=h_1=1$. Under this restriction, the condition (\ref{KTRicci}) 
is fulfilled, which in turn guarantees the separability of the Klein-Gordon equation~\cite{Galtsov:2025nia}. 
A straightforward computation then shows that the Ricci tensor components 
satisfy
\begin{align}
\label{CCLPRicci}
R_{23}=0\,, \qquad 
\frac{R_{00}+R_{22}}{Q} +\frac{R_{11}-R_{33}}{P}+\frac{2R_{01}}{\sqrt{QP}}=0\,.
\end{align}
As  recently clarified in \cite{Galtsov:2025nia} for the $F^1=0$ case, 
the solution within this subclass can indeed be solved explicitly.
We can proceed along an almost parallel line of reasoning in what follows.
Einstein's equations and (\ref{CCLPRicci}) imply that the scalar field $z$ fulfills 
\begin{align}
\label{}
\partial_w z \partial_w \bar z=0 \,, \qquad \partial_{\bar w} z \partial_{\bar w} \bar z= 0\,,
\end{align}
where $w=q+i p$. 
We therefore get either $\partial_w z=0$ or $\partial_{\bar w} z=0$.
Without loss of generality, we set $z=z(w)$, 
i.e., $z$ is a holomorphic function of $w$. 
It is well known that the class of holomorphic, single-valued functions on the entire complex plane with only a simple pole corresponds precisely to M\"obius transformations, that is, a linear fractional transformation of the form
\begin{align}
\label{}
z= \frac{\gamma_1 w+\gamma_2}{\gamma_3 w+\gamma_4} \,, \qquad \gamma_1\gamma_4-\gamma_2\gamma_3\ne 0 \,, 
\end{align}
where $\gamma_1$, $\gamma_2$, $\gamma_3$, $\gamma_4$ are complex constants. 
This specifies the action of ${\rm SL}(2,\mathbb C)$ on the Riemann sphere.
By requiring the spacetime is asymptotically AdS ($\phi=\chi=0$) as $q\to \infty$, we have 
$\gamma_1=\gamma_3=1$. By the shift of $q$ and $p$, one can set 
$\gamma_2=-\gamma_4=q_0+ i p_0$, where $q_0$ and $p_0$ are real constants related to scalar charges. 
Equation $R_{01}=0$ then implies $f_2''(q)+h_2''(p)=0$ [see (\ref{R01g})], which is combined with the ${\rm GL}(2,\mathbb R) $ transformation (\ref{GL2R}) to give $f_2(q)=q^2+\mathtt b_1 q-q_0^2$ and 
$h_2(p)=-p^2+\mathtt b_2 p+\mathtt b_3$, where $\mathtt b_1$,
$\mathtt b_2$, $\mathtt b_3$ are constants. 
Substituting these into $E_{00}+E_{22}=0$, one obtains 
$\mathtt b_1=\mathtt b_2=0$ and  $\mathtt b_3=p_0^2$, while 
equation $E_{00}-E_{11}=0$ yields decoupled equations
\begin{align}
\label{}
Q''(q)-4g^2 (3q^2-q_0^2)=2\mathtt c_1 \,, \qquad
P''(p)-4g^2(3p^2-p_0^2)=-2\mathtt c_1 \,,
\end{align}
where $\mathtt c_1$ is a separation constant. 
Using the rest of Einstein's equations, we obtain
\begin{align}
\label{QPCCLP}
Q(q)=\mathtt c_2+(q^2-q_0^2)[\mathtt c_1+g^2(q^2-q_0^2)] \,, \qquad 
P(p)=\mathtt c_2+(p_0^2-p^2)[\mathtt c_1+g^2 (p_0^2-p^2)] \,,
\end{align}
where $\mathtt c_2$ is an integration constant. 
This completes the process of solving the background field equations.
The background solution then takes the form
\begin{subequations}
\label{CCLPbackground}
\begin{align}
\D s^2 =&\,  -\frac{Q(q)}{W}(\D \tau-(p^2-p_0^2) \D \sigma)^2+\frac{P(p)}{W}(\D \tau+(q^2-q_0^2) \D \sigma)^2+W \left(
\frac{\D q^2}{Q(q)}+\frac{\D p^2}{P(p)}\right)\,, \label{CCLPbackgroundmetric}\\
z=&\, \frac{q+q_0+i(p+p_0)}{q-q_0+i(p-p_0)}\,,
\label{CCLPbackgroundscalar}
\end{align}
\end{subequations}
where $W=q^2-q_0^2+p^2-p_0^2$. 
The solution thereby generates the desired off-diagonal terms 
$f_2(q)=q^2-q_0^2$, $h_2(p)=p_0^2-p^2$, for which the scalar charges $q_0$ and $p_0$ do 
not appear in the Carter case--and thus are also absent in pure AdS.

Taking the above solution [metric (\ref{CCLPbackground}) with (\ref{QPCCLP})] as a seed, let us try to solve the full system. 
Requiring the Kerr-Schild deformed solution to be circular, 
the formulation in the above subsection implies that it suffices 
to replace $Q(q)$ by $\ti Q(q)$, while other functions remain unchanged. 
Hence, our main concern here is the configuration of the electromagnetic fields, 
which we make the following ansatz
\begin{subequations}
\label{CCLPgauge}
\begin{align}
A^0=&\,\frac{\mathtt A_{00}(q )}{W}(\D \tau-(p^2-p_0^2) \D \sigma)+
\frac{\mathtt A_{01}(p)}{W}(\D \tau+(q^2-q_0^2) \D \sigma)\,, \\
A^1=&\,\frac{\mathtt A_{10}(q )}{W}(\D \tau-(p^2-p_0^2) \D \sigma)+
\frac{\mathtt A_{11}(p)}{W}(\D \tau+(q^2-q_0^2)  \D \sigma)\,.
\end{align}
\end{subequations}
This ansatz is consistent with Ricci circularity (\ref{Riccicircular}), 
and implies  
\begin{align}
\label{}
k^\rho H_{I\rho [\mu}k_{\nu]}=0\,, \qquad 
T^{(F)} _{23}=T^{(F)} _{01}={T^{(F)} _{00}+T^{(F)} _{22}}
={T^{(F)} _{11}-T^{(F)} _{33}}=0\,.
\end{align}
It thus follows that under the Kerr-Schild deformation, one can 
continue to employ the solution to  $z$ as the background value. 
 To be specific, we need to solve the following linearized Einstein equations
\begin{align}
\label{SUGRAlinEineq}
{}^{(1)\!}R^\mu{}_\nu=T^{(F)\mu}{}_\nu \,. 
\end{align}
With this choice made, 
the Maxwell equation $\D H_I=0$ leads to 
\begin{align}
\label{}
\mathtt A_{00}(q )=\mathtt Q_0 (q-q_0)\,, \qquad 
\mathtt A_{01}(q )=\mathtt P_0 (p-p_0)\,,\qquad 
\mathtt A_{10}(q )=\mathtt Q_1 (q+q_0)\,, \qquad 
\mathtt A_{11}(q )=\mathtt P_1 (p+p_0)\,,
\end{align}
where $\mathtt Q_0$, $\mathtt Q_1$
$\mathtt P_0$, $\mathtt P_1$ are constants, related to the 
electric and magnetic charges. 
Note that the Maxwell equation is insensitive to $\ti Q$,
or equivalently to the deformation function $H$. 

Substituting the above expressions into linearized Einstein's equations (\ref{SUGRAlinEineq})
with (\ref{solH}), 
one finds that the general solution takes the form
\begin{align}
\label{dyonicQ}
\ti Q(q)=\mathtt a_2+(q^2-q_0^2)[\mathtt a_1+g^2(q^2-q_0^2)]+\mathtt a_3 q \,,
\end{align}
where the constants $\mathtt a_1, \mathtt a_2, \mathtt a_3$ are restricted by
\begin{align}
\label{}
\mathtt a_1=\mathtt c_1\,, \qquad 
\mathtt a_2 =\mathtt c_2+\frac 12 (\mathtt P_0^2+\mathtt P_1^2+\mathtt Q_0^2
+\mathtt Q_1^2)\,, \qquad 
\mathtt a_3= \frac{\mathtt P_0^2-\mathtt P_1^2-\mathtt Q_0^2+\mathtt Q_1^2}{2q_0}
\,,
\end{align}
together with an additional constraint
\begin{align}
\label{}
p_0 (\mathtt P_0^2-\mathtt P_1^2-\mathtt Q_0^2+\mathtt Q_1^2)
+2 q_0 (\mathtt P_0\mathtt  Q_0-\mathtt P_1\mathtt Q_1) =0 \,. 
\end{align}
The scalar field equation (\ref{N2zeq}) is then automatically satisfied. 

Thus, we have obtained a new dyonic solution 
[metric (\ref{CCLPbackgroundmetric}) with the replacement $Q\to \ti Q$, 
scalar field (\ref{CCLPbackgroundscalar}) and gauge fields (\ref{CCLPgauge})]
with nonvanishing $p_0$ through the Kerr-Schild formalism. 
For the $F^1=0$ case, the solution recovers the one in \cite{Wu:2020mby}. 
As we have sen, present construction based on the Kerr-Schild ansatz becomes feasible only when one starts from a sufficiently general background solution (\ref{CCLPbackground}), which accommodates the required off-diagonal structure. This in turn illustrates the versatility of the present algorithm, showing its capability to generate solutions that would otherwise be inaccessible within more restrictive frameworks.

\subsection{Conformal to degenerate BF metric}
\label{sec:conformalBF}

We next consider a class of spacetimes whose metric is conformal to  the degenerate BF metric as 
\begin{align}
\label{}
\D s^2 =\Omega(p,q) ^2 [S_1(q)+S_2(p)] \Biggl[
-\frac{Q(q)}{W^2}\left(h_1(p)\D \tau+h_2(p) \D \sigma \right)^2
+\frac{P(p)}{W^2}\left(f_1(q)\D \tau+f_2 (q)\D \sigma\right)^2
+\frac{\D q^2}{Q(q)}+\frac{\D p^2}{P(p)}
\Biggr]\,,
\end{align}
where $\Omega(p,q) $ denotes the conformal factor\footnote{Althogh the overall
factor $S_1(q)+S_2(p)$ can be absorbed into $\Omega$, we retain it for easier comparison with the Carter class.} and $W$ is defined as (\ref{W}). 
This metric admits two commuting Killing vectors $\partial/\partial \tau$ and $\partial/\partial \sigma$
and satisfies the circularity condition. In contrast to the original degenerate BF metric, it does not 
admit a Killing tensor. Despite this, the null geodesics remain separable, since they are insensitive to the conformal factor.
In particular,  the tangent vector (\ref{null}) corresponds to the 
shear-free null geodesics maintaining the constant angular coordinate $p$.

We now consider a Kerr-Schild-type deformation of the above conformally degenerate BF metric, as defined in (\ref{KerrSchild}), and impose the condition that the deformed metric still belongs to the class of circular spacetimes. Under this requirement, the deformation function $H$ takes the form
\begin{align}
\label{solHconf}
H(p, q) = \frac{\Omega(p,q)^2[Q(q)- \ti Q(q)][S_1(q)+S_2(p)]}{2 W(p,q)^2 } \,,
\end{align}
where $\ti Q(q)$ is an arbitrary function of $q$.
By the coordinate transformation (\ref{newtausigma}), one can verify that the 
deformed metric is also conformal to the degenerate BF metric with a modified structure function $\ti Q(q)$, i.e., 
\begin{align}
\label{}
\D \ti s^2 =&\,\Omega(p,q) ^2 [S_1(q)+S_2(p)] \Biggl[
-\frac{\ti Q(q)}{W^2}\left(h_1(p)\D \ti \tau+h_2(p) \D \ti \sigma \right)^2
\notag \\ &\,
+\frac{P(p)}{W^2}\left(f_1(q)\D \ti \tau+f_2 (q)\D \ti \sigma\right)^2+\frac{\D q^2}{\ti Q(q)}+\frac{\D p^2}{P(p)}
\Biggr]\,. 
\end{align}

This incorporates the  case of Pleba\'nski-Demia\'nski family of solution~\cite{Plebanski:1976gy},
for which the metric is conformal to Carter family (\ref{Carter}) with $\Omega=1/(1-pq)$. 
By choosing $Q$ and $P$ to be suitable quartic functions, 
this metric describes the most general Petrov-D spacetime for vacuum Einstein's equations with 
a cosmological constant. See e.g., \cite{Griffiths:2005qp,Klemm:2013eca,Nozawa:2015qea,Astorino:2023elf}
for the physical properties of the Pleba\'nski-Demia\'nski  solution.

\section{$D=5 $}
\label{sec:5D}

The BF metric in $D=4$ is distinguished by two mutually commuting Killing vectors together with a nontrivial Killing tensor. These symmetries underlie the integrability of the geodesic equations and the separability of various field equations in this background. 
A natural question then arises as to whether this construction can be extended to higher dimensions. In fact, a five-dimensional generalization is possible, and the corresponding metric exhibits an even richer set of symmetries, admitting three mutually commuting Killing vectors in addition to a Killing tensor. This extension preserves the essential features of the BF class while revealing new geometric structures specific to higher dimensions. The development and analysis of this $D=5$ generalization constitute the central subject of the present section.

\subsection{Degenerate BF metric}

Let us consider the $D=5$ dimensional BF metric (\ref{Dmetric}), where the Killing coordinates are denoted by $\psi^i = (\tau, \sigma, \chi)$ and are now regarded as spanning a three-dimensional subspace. The associated Killing vector fields are given by
\begin{align}
\label{KVs5D}
\xi_{(1)}=\frac{\partial}{\partial \tau}\,, \qquad 
\xi_{(2)}=\frac{\partial}{\partial \sigma}\,, \qquad 
\xi_{(3)}=\frac{\partial}{\partial \chi}\,.
\end{align}
We impose  that both of the matrices $F_1^{ij}(q)$ and $F_2^{ij}(p)$ have rank two:
\begin{align}
\label{}
{\rm rank}F_1^{ij}={\rm rank}F_2^{ij}=2 \,. 
\end{align}
Writing Hamilton's principal function as (\ref{HamiltonS}), 
we require that the metric admits special null geodesics along which $p={\rm const.}$, i.e., 
\begin{align}
\label{PND5D}
\frac{\D \ma S_p}{\D p}=\ma C=m=0 \,. 
\end{align}
where $\ma C$ is a separation constant between $q$ and $p$. 
We further assume that the constant $\ma P_\chi=p_\mu (\partial/\partial\chi)^\mu$ of the geodesic motion  vanishes
\begin{align}
\label{Pchi}
\ma P_\chi=0 \,.  
\end{align} 
This imposes a distinct additional condition on each of the matrices $F_1^{ij}(q)$ and $F_2^{ij}(p)$. 
As a consequence, each matrix $F_1^{ij}(q)$ and $F_2^{ij}(p)$ has four independent components.  
Then, these matrices are parametrized as
\begin{align}
\label{}
F_1^{ij}(q)=&\, -\frac 1{Q(q)} \left(
\begin{array}{ccc}
f_2^2(q)  & -f_1(q)f_2(q)  & -f_2(q)f_3(q)    \\
-f_1(q)f_2(q) & f_1^2(q) &  f_1(q)f_3(q)\\
-f_2(q)f_3(q) &   f_1(q)f_3(q) &  f_3^2(q)- f_0(q) Q(q)
\end{array}
\right)  \,, 
\\
F_2^{ij}(p)=&\, -\frac 1{P(p)}
\left(
\begin{array}{ccc}
  h_2^2(p)  &-h_1(p) h_2(p) &- h_2(p)h_3(p)   \\
  -h_1(p)h_2(p)& h_1^2(p) & h_1(p)h_3(p) \\ 
 -h_2(p)h_3(p)   & h_1(p)   h_3(p) & h_3^2(p)-h_0(p)P(p) 
\end{array}
\right)\,.
\end{align}
Then, the corresponding five-dimensional metric is expressed as 
\begin{align}
\label{}
\D s^2 = \,& \left[S_1(q)+S_2(p)\right] \Biggl[
-\frac{Q(q)}{W_3(p,q)^2}\left(h_1(p)\D \tau +h_2(p) \D \sigma \right)^2
+\frac{P(p)}{W_3(p,q)^2}\left(f_1(q)\D \tau +f_2(q) \D \sigma \right)^2
\notag \\
&+\frac{\D q^2}{Q(q)}+\frac{\D p^2}{P(p)}+ \frac{1}{W_3(p,q)^2 W_0(p,q)}\left(
W_3(p,q) \D \chi -W_2(p,q) \D \tau+W_1(p,q) \D \sigma 
\right)^2 
\Biggr]\,, 
\label{5Dmetric}
\end{align}
where we have defined 
\begin{subequations}
\begin{align}
\label{}
W_1(q,p)=&\,f_3(q)h_2(p)-f_2(q)h_3(p) \,, \\
W_2(q,p)=&\,f_1(q)h_3(p)-f_3(q)h_1(p) \,, \\
W_3(q,p)=&\,f_2(q)h_1(p)-f_1(q)h_2(p)\,, \\
W_0(q,p)=&\, f_0(q)-h_0(p) \,.
\end{align}
\end{subequations}
We shall also refer to the $D=5$ metric (\ref{5Dmetric}) as the degenerate BF metric, 
which is now specified by twelve functions.

The associated null vectors satisfying 
(\ref{PND5D}) and (\ref{Pchi}) are given by
\begin{align}
\label{5Dvec}
k_\mu ^{(\pm)}\D x^\mu =-h_1(p) \D \tau-h_2(p)\D \sigma \pm \frac{W_3(p,q)}{Q(q)}\D q \,. 
\end{align}
It should be emphasized that, in $D=5$, the above null vector along which the angular 
coordinate $p$ remains constant does not share the same additional properties as the four-dimensional case. In particular, it is neither shear-free nor does it coincide with the principal null directions  (\ref{Weyleig}). Furthermore, it does not necessarily correspond to the Weyl aligned null direction
$C_{\mu\rho\nu\sigma}k^\rho k^\sigma=0$. 
This distinction highlights an important limitation of the direct analogy with four-dimensional spacetimes: while in 
$D=4$ the existence of a shear-free principal null congruence aligned with the Weyl tensor plays a crucial role in the Petrov classification and in the separability of field equations, such a privileged structure no longer persists in higher dimensions.
See e.g., \cite{Frolov:2003en,Coley:2004jv} for details.

The above degenerate BF metric (\ref{5Dmetric}) shares the invariance properties similar to the four dimensional case. 
The shift and scaling symmetry of $S_1$ and $S_2$ now reads 
\begin{align}
\label{shift0}
S_1 \to \mathsf s_0 S_1+\mathsf s_1 \,, \qquad 
S_2 \to  \mathsf s_0 S_2-\mathsf s_1 \,, \qquad 
f_0 (q)\to \mathsf s_0^{-1}f_0(q)+\mathsf s_2 \,, \qquad 
h_0 (p)\to \mathsf s_0^{-1}h_0(p)+\mathsf s_2 \,,
\end{align}
accompanied by 
\begin{align}
\label{}
Q\to \mathsf s_0 Q\,, \qquad 
P \to \mathsf s_0 P\,, \qquad 
\tau \to \mathsf s_0^{-1} \tau \,, \qquad 
\sigma\to \mathsf s_0^{-1} \sigma \,, 
\qquad 
\chi \to \mathsf s_0^{-1} \chi \,. 
\end{align}
Under the coordinate reparametrization 
$q\to \check q(q)$ and $p\to \check p(p)$, 
the metric is also invariant provided $f_i$ and $h_i$ ($i=1,2,3$) transform according to (\ref{fhitr}), whereas
$f_0$ and $h_0$ remain inert. 
Furthermore, there exists the following freedom contained in ${\rm GL}(3,\mathbb R)$ corresponding to the 
linear mixing of Killing coordinates $\psi^i=(\tau, \sigma, \chi)$
\begin{align}
\label{5Daij}
\psi^i \to \mathsf A^i{}_j \psi^j \,, \qquad 
\mathsf A^i{}_j=\left(
\begin{array}{ccc}
\mathsf  a_{11} & \mathsf a_{12} & 0    \\
\mathsf  a_{21} &  \mathsf a_{22} & 0 \\ 
 \mathsf a_{31} &\mathsf  a_{32} & \mathsf a_{33}   
\end{array}
\right) \,,
\end{align}
with the redefinition 
\begin{align}
\label{}
f_i (q) \to \mathsf B_i{}^j f_i(q) \,, \qquad 
 h_i (p) \to \mathsf B_i{}^j h_i(p) \,, \qquad 
 f_0(q)\to \mathsf a_{33}^2 f_0(q) \,, \qquad 
 h_0(p)\to \mathsf a_{33}^2 h_0(p)\,,
\end{align}
where $i, j=1,2,3 $ and 
\begin{align}
\label{}
\mathsf  B_i{}^j=\left(
\begin{array}{ccc}
\mathsf a_{22}  & -\mathsf a_{21}  &  0 \\
-\mathsf a_{12} &   \mathsf a_{11} & 0 \\ 
\mathsf a_{32} & -\mathsf a_{31} & \mathsf a_{33} 
\end{array}
\right)\,. 
\end{align}
Here $\mathsf a_{ij}$ are constants with ${\rm det}(\mathsf a_{ij})\ne 0$. 

As a notable example, 
the metric (\ref{5Dmetric}) recovers the Chen-L\"u-Pope family \cite{Chen:2006xh} corresponding to the 
four-dimensional Carter family,  by the following choice
\begin{align}
\label{5DCLPfh}
f_1(q)=&\, h_1(p)=1\,, & S_1(q)=&\,f_2(q)= q^2\,, & S_2(p)=&\, -h_2(p)=p^2 \,, \notag \\
f_3(q)=&\, a_0 f_0 (q)=q^{-2} \,, & h_3(p)=&\, a_0 h_0(p)=-p^{-2} \,,
\end{align}
where $a_0$ is a constant. 
The metric is then simplified to 
\begin{align}
\label{}
\D s^2 =&\,
-\frac{Q(q)}{q^2+p^2}(\D \tau-p^2 \D \sigma)^2 
+\frac{P(p)}{q^2+p^2}(\D \tau+q^2 \D \sigma)^2 +(q^2+p^2)\left(\frac{\D q^2}{Q(q)}+\frac{\D p^2}{P(p)}\right)\notag \\
&+\frac{a_0}{p^2q^2}\left(\D \tau+(q^2-p^2)\D \sigma+p^2 q^2 \D \chi \right)^2\,.
\label{CLP}
\end{align}
For comparison with the notation in~\cite{Chen:2006xh}, set $x=q^2$, $y=p^2$, 
$X(x)=q^2 Q$ and $Y(y)=p^2 P$. 
As we will demonstrate in the appendix, only 
the Chen-L\"u-Pope family (\ref{CLP}) admits an Einstein metric 
within the class of degenerate BF metrics (\ref{5Dmetric}). This solution possesses a Killing-Yano tensor \cite{Houri:2007uq} and contains the doubly rotating Myers-Perry-(A)dS metrics  as a special case.

\subsection{Kerr-Schild transformation}

Next, we consider the Kerr-Schild deformation of the metric with $k_\mu=k_\mu^{(+)}$
as (\ref{KerrSchild}). 
We require that the deformed metric still admits 
the Killing vectors (\ref{KVs5D}) and remains circular (\ref{circularD}). 
These  conditions uniquely  determine the possible form of $H=H(p,q)$ as
\begin{align}
\label{}
H(p,q)=\frac{[Q(q)-\ti Q(q)][S_1(q)+S_2(p)]}{2W_3(p,q)^2}\,,
\end{align}
where $\ti Q=\ti Q(q)$ is an arbitrary function of $q$. 
Transforming to the new coordinates 
\begin{subequations}
\begin{align}
\label{}
\ti \tau=&\,\tau+\int f_2(q) \left(\frac{1}{\ti Q(q)}-\frac 1{Q(q)}\right)\D q \,, \\
\ti \sigma =&\, \sigma-\int f_1(q)  \left(\frac{1}{\ti Q(q)}-\frac 1{Q(q)}\right)\D q \,, \\
\ti \chi =&\, \chi-\int f_3(q) \left(\frac{1}{\ti Q(q)}-\frac 1{Q(q)}\right)\D q \,, 
\end{align}
\label{5Dcoordtr}
\end{subequations}
the deformed metric is simplified to 
\begin{align}
\label{}
\D \ti s^2 = \,& \left[S_1(q)+S_2(p)\right] \Biggl[
-\frac{\ti Q(q)}{W_3(p,q)^2}\left(h_1(p)\D \ti \tau +h_2(p) \D \ti \sigma \right)^2
+\frac{P(p)}{W_3(p,q)^2}\left(f_1(q)\D \ti \tau +f_2(q) \D \ti \sigma \right)^2
\notag \\
&+\frac{\D q^2}{\ti Q(q)}+\frac{\D p^2}{P(p)}+ \frac{1}{W_3(p,q)^2 W_0(p,q)}\left(
W_3(p,q) \D \ti \chi -W_2(p,q) \D \ti \tau+W_1(p,q) \D \ti \sigma 
\right)^2 
\Biggr]\,. 
\end{align}
It follows that the deformed geometry also belongs to the family of degenerate BF metrics, analogous to the situation encountered in the four-dimensional case. The only difference is the specific form of the structure function, which changes as $Q\to \ti Q$.

This class of metrics includes the asymptotically AdS  charged rotating black hole in minimal gauged supergravity
\cite{Chong:2005hr}, which is recovered by the choice (\ref{5DCLPfh}) except for 
$f_3(q) =(1+\ma Q_e)/q^2 $, where $\ma Q_e$ is a constant related to the electric charge (see \cite{Lu:2008ze}). 
A direct consequence of this modification is that the resulting geometry cannot be expressed within the Kerr-Schild ansatz built upon the AdS background, since it no longer belongs to the Chen-L\"u-Pope class of Einstein spaces.
In our formulation this conclusion is immediate, while the analysis in \cite{Malek:2014dta} requires a considerable amount of computational effort. This highlights the efficiency and transparency of our approach.

\subsection{Non-conformal distortion of the degenerate BF metric}

We next explore the distortion of the $D=5$ degenerate BF metric (\ref{5Dmetric}). 
As in the four dimensional case, the overall conformal distortion is obvious, since 
the null geodesics are unaffected by conformal transformation. 
More nontrivially,  the following class of nonconformal distortion is possible 
\begin{align}
\label{}
\D s^2 = \,& \Omega_1(p,q)^2 \left[S_1(q)+S_2(p)\right] \Biggl[
-\frac{Q(q)}{W_3(p,q)^2}\left(h_1(p)\D \tau +h_2(p) \D \sigma \right)^2
+\frac{P(p)}{W_3(p,q)^2}\left(f_1(q)\D \tau +f_2(q) \D \sigma \right)^2
\notag \\
&+\frac{\D q^2}{Q(q)}+\frac{\D p^2}{P(p)}\Biggr]+ \frac{\Omega_2(p,q)^2 \left[S_1(q)+S_2(p)\right]}{W_3(p,q)^2 W_0(p,q)}\left(
W_3(p,q) \D \chi -W_2(p,q) \D \tau+W_1(p,q) \D \sigma 
\right)^2 
\,, 
\label{5Ddefmetric}
\end{align}
where $\Omega_1(p,q)$ and $\Omega_2(p,q)$ represent independent distortion factors depending solely 
on the coordinates $p$ and $q$.
Even in this case, 
one can explicitly verify that the null vector (\ref{5Dvec}) continues to satisfy the geodesic equation for the distorted metric (\ref{5Ddefmetric}), along which the angular coordinate is unchanged ($p={\rm const.}$). 
The underlying reason for this persistence lies in the vanishing of the conserved momentum $\ma P_\chi =0$,
which guarantees that the congruence remains geodesic despite the anisotropic distortion introduced by unequal scaling factors.
The $\Omega_1\ne \Omega_2$ case therefore encompasses a considerably broader family of spacetimes than the conformally rescaled subclass, opening the possibility of exploring richer geometric and physical phenomena within the framework of five-dimensional Kerr-Schild-type constructions.

Indeed, this generalized framework includes, as a special case, the class of vacuum solutions constructed by L\"u, Mei, and Pope~\cite{Lu:2008js,Lu:2008ze}.  Specifically, the L\"u-Mei-Pope solution is obtained by 
$\Omega_1=1/(1-p^2q^2)$, $\Omega_2=1$ and (\ref{5DCLPfh}), viz, 
\begin{align}
\label{}
\D s^2 =&\, \frac 1{(1-p^2q^2)^2}\Biggl[
-\frac{Q(q)}{q^2+p^2}(\D \tau-p^2 \D \sigma)^2 
+\frac{P(p)}{q^2+p^2}(\D \tau+q^2 \D \sigma)^2 +(q^2+p^2)\left(\frac{\D q^2}{Q(q)}+\frac{\D p^2}{P(p)}\right)\Biggr]\notag \\
&+\frac{a_0}{p^2q^2}\left(\D \tau+(q^2-p^2)\D \sigma+p^2 q^2 \D \chi \right)^2\,,
\label{LMP}
\end{align}
with the structure functions
\begin{align}
\label{QPLMP}
Q(q)=a_0 q^{-2}+a_1+a_2 q^2+a_3 q^4 +a_0 q^6 \,, \qquad 
P(p)= -\left(a_0 p^{-2}+a_3 +a_2 p^2+a_1 p^4+a_0 p^6\right) \,. 
\end{align}
Here, $a_0$, $a_1$, $a_2$, $a_3$ are constants. 
With suitable choice of parameters, this solution describes a black hole with a lens space topology $L(m,n)$ in an asymptotically locally flat space with cross-section topology $L(n,m)$. Moreover, the above metric (\ref{LMP}) encompasses the Emparan-Reall black ring $S^1 \times S^2$~\cite{Emparan:2001wn} under the double Wick rotation.\footnote{Note that the condition $\ma P_\chi = 0$ corresponds to null geodesics with zero energy in the black ring spacetime.}

Let us consider the Kerr-Schild deformation $\ti g_{\mu\nu}=g_{\mu\nu}+2Hk_\mu k_\nu$ of the metric 
(\ref{5Ddefmetric}) and require that it preserve both the Killing symmetries and the circularity condition.
In this case, the scalar function $H = H(p,q)$ is found to take the form
\begin{align}
\label{H_LMP}
 H(p,q)=\frac{\Omega_1(p,q)^2 [Q(q)-\ti Q(q)][S_1(q)+S_2(p)]}{2W_3(p,q)^2}\,.
 \end{align}
Proceeding in a manner analogous to the undeformed case, one finds that the coordinate transformation given in (\ref{5Dcoordtr}) brings the deformed metric $\D \tilde s^2=\ti g_{\mu\nu}\D \ti x^\mu \D \ti x^\nu $ back into the canonical form (\ref{5Ddefmetric}), now with the replacement $Q(q) \to \tilde Q(q)$. This demonstrates the consistency of the deformation procedure within the class of metrics admitting separable null geodesics.

Although the writing of the distorted metric (\ref{5Ddefmetric}) into the Kerr-Schild ansatz (\ref{KerrSchild}), (\ref{5Ddefmetric}), 
(\ref{H_LMP}) is highly suggestive, the mere $Q\to \ti Q$ prescription from a simpler seed solution does not work, 
as far as the L\"u-Mei-Pope vacuum solution (\ref{LMP}) is concerned. 
As evident from the expressions in (\ref{QPLMP}), 
the functional form of $Q$ and $P$ for the L\"u-Mei-Pope solution is restricted by $Q(q)=-q^4 P(1/q)$. 
Hence, altering only $Q$ does not work, and the solution to ${}^{(1)\!}R^\mu{}_\nu=0$ 
becomes trivial $H=0$. 
This outcome is fully consistent with the general analysis in \cite{Ortaggio:2008iq,Srinivasan:2025hro}.
To see this, let us introduce the frame $\{k_\mu, n_\nu, m_{(i)\mu}\}$ in $D$ dimensions by 
$g_{\mu\nu}=-2 k_{(\mu}n_{\nu)}+\delta_{ij}m_{(i)\mu} m_{(j)\nu}$, where 
\begin{align}
\label{}
k^\mu n_\mu=-1 \,, \qquad m_{(i)\mu}m_{(j)}{}^\mu=\delta _{ij} \,, \qquad k^\mu m_{(i)\mu}=n^\mu m_{(i)\mu}=0 \,,
\end{align}
By (\ref{RicciKS}), we obtain 
\begin{align}
\label{Rijdec}
\ti R_{\mu\nu}m^\mu_{(i)}m^\nu_{(i)}&=R_{\mu\nu}m^\mu_{(i)}m^\nu_{(i)}-2\left\{HR_{\mu\rho\nu\sigma}m^\mu_{(i)}m^\nu_{(i)}k^\rho k^\sigma
+ H L_{ki}L_{kj}-[\ma DH+(D-2)\theta H]S_{ij}\right\}\,,
\end{align}
where we have denoted $\ma DH=k^\mu \nabla_\mu H$, $L_{ij}=m^\mu_{(i)}m^\nu _{(j)}\nabla_\mu k_\nu $, $S_{ij}=L_{(ij)}$ and 
$\theta =S_{ii}/(D-2)$. 
Imposing $ R_{\mu\nu}=0$ and $\ti R_{\mu\nu}=0$, the terms in curly brackets in 
(\ref{Rijdec}) vanish. Taking the trace of this equation and eliminating $\ma D H$ (noting that $\theta \ne 0$ in the present case), one obtains  
the (generalized) optical constraint \cite{Ortaggio:2008iq,Srinivasan:2025hro}
\begin{align}
\label{GOC}
R_{\mu\rho\nu\sigma}m^\mu_{(i)}m^\nu_{(i)}k^\rho k^\sigma=-L_{ki}L_{kj}+\frac 1{(D-2)\theta}L_{kl}L^{kl}S_{ij}
\end{align}
In the analysis of \cite{Ortaggio:2008iq,Srinivasan:2025hro}, which concerns Weyl aligned null directions--including the $D=5$ Myers-Perry-AdS metric--the left-hand side of (\ref{GOC}) vanishes. 
However, a direct verification confirms that the null vector (\ref{5Dvec}) for the L\"u-Mei-Pope solution (\ref{LMP}) does not obey this condition, thereby clarifying why it cannot be generated through such a straightforward prescription.

Accordingly, in order for the present formulation to succeed, it is necessary to consider nonvacuum configurations arising from (non)vacuum seeds. A detailed analysis of this case is left for future work.

\section{Concluding remarks}
\label{sec:conclusion}

In this paper, we have investigated the Kerr-Schild formalism in the context of the BF family of metrics. Our analysis has been directed toward the degenerate subclass (\ref{metric}), together with the shear-free null geodesics defined in (\ref{null}), under the additional requirement that the deformed geometry retains the Killing symmetry and circularity. Quite remarkably, we find that any such Kerr-Schild deformation still falls into the degenerate BF family. In fact, the transformation reduces simply to replacing the structure function $Q(q)$ by a new one $\tilde Q(q)$, while leaving the other form of the metric components intact.

This observation provides an organizing framework for the Kerr-Schild transformation within the BF class. The importance of this result lies in the fact that it preserves the hidden symmetries and integrability properties characteristic of the BF metrics. In particular, the separability of the geodesic equations in the new geometry is directly inherited from the seed solution. To the best of our knowledge, such a systematic way of generating Kerr-Schild deformations within the BF family has not been explicitly recognized in the literature.

As an illustrative example, we have considered the rotating black hole solution of the ${\cal N}=2$ gauged supergravity. By starting from a solution of Einstein-scalar gravity with a scalar potential, rather than pure AdS, we were able to generate a new dyonic configuration (\ref{dyonicQ}). This outcome highlights the effectiveness of the present algorithm.

We have also considered a generalization to five dimensions by assuming that the null geodesics admit a fixed angular coordinate, with one of the constants of geodesic motion set to vanish $\ma P_\chi = 0$. This class of metrics (\ref{5Dmetric})
 includes the doubly rotating Myers-Perry-AdS black hole. In $D=5$, the metric can be distorted in such a way that two distinct distortion factors appear (\ref{5Ddefmetric}), which does not occur in $D=4$. Under the Kerr-Schild transformation, this class of solutions can likewise be cast into the same family, but with a modified structure function $\tilde Q$. Although this prescription does not work for the L\"u-Mei-Pope solution (\ref{LMP}), it  opens a new avenue for enlarging the landscape of exact solutions in general relativity. 

For definiteness of the argument, this paper restricts attention to the cohomogeneity-two case in which the $D$-dimensional metric possesses the $D-2$ commuting Killing vectors with a nontrivial Killing tensor. For $D\ge 6$, this class does not encompass the Myers-Perry-(A)dS metrics. The extension to the broader BF class with $D-n$ Killing vectors and $n$ Killing tensors would be worthwhile to pursue. In particular, it would be interesting to see whether this broader framework can reproduce and further generalize the results obtained in~\cite{Deshpande:2024vbn}.

A further direction is to seek generalizations of the Pleba\'nski-Demia\'nski solution with a scalar potential. Also, rotating generalizations of hairy black holes~\cite{Faedo:2015jqa} and wormholes~\cite{Nozawa:2020gzz, Nozawa:2023aep} in AdS provide promising avenues  for future investigation and are currently being pursued.

Another natural extension is to consider metrics lying outside the degenerate BF class. For instance, the solutions of \cite{Wu:2011zzh} in $D=4$ (c.f also \cite{Wu:2011gq}) are conformal to the nondegenerate BF family. When written in Kerr-Schild form with respect to the AdS background, the associated deformation vector is not null, and a nonconformal distortion becomes necessary. Revisiting such cases within a generalization of the present framework would be worthy of further investigation.

\section*{Acknowledgments}
We would like to thank Tsuyoshi Houri for bringing some relevant references to our attention. 
The work of MN is partially supported by MEXT KAKENHI Grant-in-Aid for Transformative Research Areas (A) through the ``Extreme Universe'' collaboration JP21H05189 and JSPS KAKENHI Grant Number JP20K03929, JP25K07309. 
The work of TT is supported by JSPS KAKENHI Grant Number JP22K18604, JP24K21170, JP24K07029, JP25K07290.

\appendix

\section{Classification of Einstein space}
\label{sec:Einstein}

In this appendix, we explore the condition under which the 
degenerate BF class is Einstein $R_{\mu\nu}=\frac{2}{D-2}\Lambda g_{\mu\nu}$, 
where $\Lambda $ is a cosmological constant.

\subsection{$D=4$}

In the $D=4$ case, the Goldberg-Sachs theorem \cite{GS} enables us to conclude that the 
metric must belong to the Petrov-D class. It has been known that the most general Petrov-D Einstein metric is 
the Pleba\'nski-Demia\'nski family \cite{Plebanski:1976gy}. Among this class, the separable timelike geodesics
are achieved only in the vanishing acceleration limit, which reduces to the Carter class. 
This result has been already obtained in the literature~\cite{Anabalon:2016hxg}, 
but we shall discuss the detail for comparison to the $D=5$ case.

By inspecting (\ref{Rpq}), we obtain 
\begin{align}
\label{A1A2}
\frac{S_1(q)+S_2(p)}{W}=A_1(q) A_2(p)\,,   
\end{align}
where $A_1(q)$ and $A_2(p)$ are arbitrary nonvanishing functions 
of their respective arguments. 
By using the freedom $q\to\check q(q)$ and $p\to \check p(p)$ in (\ref{fhitr}), 
we can choose $f_1=1/A_1$ and $h_1=1/A_2$.
In this case, the condition (\ref{A1A2}) implies $f_2(q)=(S_1(q)+s_0)/A_1(q)$, $h_2(p)=-(S_2(p)-s_0)/A_2(p)$,
where $s_0$ is a constant. One can set $s_0=0$ by absorbing into the shift of 
$S_1$ and $S_2$. Furthermore, the redefinition $Q\to A_1^{-2} Q$, $P\to A_2^{-2} P$
and the rescaling $\D q\to \D \ti q= \D q/A_1$, $\D p\to \D \ti p= \D p/A_2$
enable us to set $A_1=A_2 = 1$ without loss of generality.

Inspecting (\ref{A1A2}) and (\ref{R01g}), we obtain 
$f_2''(q)+h_2''(p)=0$ and thus 
\begin{align}
\label{}
f_2(q)=b_0+b_1 q+b_2 q^2 \,, \qquad h_2(p)=b_3+b_4 p-b_2 p^2 \,,
\end{align}
where $b_0$, $b_1$, $b_2$, $b_3$, $b_4$ are constants. 
The following analysis can be divided into (i) $b_2 \ne 0$ and (ii) $b_2=0$.

If $b_2\ne 0$, one can set $b_2=1$ and 
$b_1=b_4 =0 $ by the rescaling and the constant shift of $q$ and $p$. 
Inserting this into $R^0{}_0+R^1{}_1=2\Lambda$, 
we have $Q''(q)+4\Lambda f_2(q)=4\Lambda h_2(p) -P''(p)$, thereby $Q$ and $P$ are quartic functions.
The rest of Einstein conditions fixes $b_3=b_0$, which can be set to $b_0=0$ 
by the ${\rm GL}(2, \mathbb R)$ transformation (\ref{GL2R}),  and 
\begin{align}
\label{CarterPQE}
Q(q)=a_0+a_1 q+a_2q^2-\frac 13 \Lambda q^4 \,, \qquad 
P(p)=a_0+a_3 p-a_2 p^2 -\frac 13 \Lambda p^4 \,,
\end{align}
where $a_0$, $a_1$, $a_2$, $a_3$ are constants. It follows that 
the metric reduces to the Carter family (\ref{Carter}). 
One recovers the maximally symmetric spacetimes by $a_1=a_3=0$.

When $b _2=0$, the Einstein condition calls for $b_1=b_4=0$. 
The freedom (\ref{GL2R}) enables us to  set $\D \tau+b_0 \D \sigma\to \D \tau$, 
$\D \tau+b_3 \D \sigma \to \D \sigma $, 
and $b_0-b_3=1$. Then, 
the metric takes the product form
\begin{align}
\label{Narial}
\D s^2=-Q_{\rm N}(q) \D \tau^2+\frac{\D q^2}{Q_{\rm N}(q)} +P_{\rm N}(p)\D \sigma^2+\frac{\D p^2}{P_{\rm N}(p)}\,,
\end{align}
where the rest of Einstein's equations fix
\begin{align}
\label{}
Q_{\rm N}(q)=a_0'+a_1'q-\Lambda q^2\,, \qquad 
P_{\rm N}(p)=a_2'+a_3'p-\Lambda p^2\,.
\end{align}
Here, $a_{0}'$, $a_{1}'$, $a_{2}'$, $a_{3}'$ are constants. 
This represents the Nariai class ${\rm dS}_2\times S^2$ or 
${\rm AdS}_2\times H^2$ \cite{Nariai:1999iok}. 
It is worth noting that  the Nariai  classes can be obtained from the Carter class by taking appropriate limiting procedures.

This concludes that the only Einstein metric contained in the degenerate BF metric is 
exhausted by the Carter family. 
The Carter family includes as a special case the Kerr-AdS metric ($\Lambda=-3g^2$).
The explicit transformation to the Boyer-Lindquist coordinates (\ref{KerrAdS}) is given by 
\begin{align}
\label{}
q=r\,, \qquad p=a \cos \theta \,, \qquad 
\tau =t-\frac{a \phi}{1-a^2g^2} \,, \qquad \sigma=-\frac \phi {a(1-a^2g^2)} \,,
\end{align}
with $Q=\Delta_r$, $P=a^2\sin^2\theta \Delta_\theta$, and 
\begin{align}
\label{}
a_0 =a^2 \,, \qquad a_1=-2M \,, \qquad a_2=1+a^2g^2 \,, \qquad 
a_3=0 \,. 
\end{align}

\subsection{$D=5$}

Next, we explore the condition under which 
the five-dimensional degenerate BF metric (\ref{5Dmetric}) is Einstein. 
Due to the inapplicability of the Goldberg-Sachs theorem, the classification performed here 
is nontrivial and appears to be original. 
Employing the frame
\begin{align}
\label{}
e^0=&\,\frac{\sqrt{Q(S_1+S_2)}}{W_3}(h_1\D \tau+h_2 \D \sigma)\,, \qquad 
e^1=\frac{\sqrt{P(S_1+S_2)}}{W_3}(f_1\D \tau+f_2 \D \sigma)\,, \notag \\
e^2=&\,\frac {\sqrt{S_1+S_2}}{W_3 \sqrt{W_0}}(W_3 \D \chi-W_2 \D \tau+W_1 \D \sigma)\,, 
\qquad 
e^3=\sqrt{\frac{S_1+S_2}{Q}}\D q \,, \qquad 
e^4=\sqrt{\frac{S_1+S_2}{P}}\D p \,, 
\end{align}
with $\eta_{ab}={\rm diag}(-1,1,1,1,1)$, 
the equation $R_{34}=0$ gives
\begin{align}
\label{Rpq5}
\partial_p \partial_q \left[\ln \left(\frac{(S_1+S_2)^3}{W_0 W_3^2}\right) \right]=0 \,. 
\end{align}
Thus, we obtain $[S_1(q)+S_2(p)]^{3/2}/(W_0^{1/2}W_3) =A_1 (q)A_2(p)$. 
By the gauge freedom (\ref{fhitr}), one can choose $f_1(q)=c_0q /A_1(q)$ and 
$h_1(p)=c_0 p/A_2(p)$, where $c_0$ is an arbitrary constant. Further redefinitions
$f_{2,3}\to  c_0 qf_{2,3} /A_1$, $h_{2,3}\to c_0p h_{2,3}/A_2$, $Q\to c_0^2 q^2 Q /A_1^2$, 
$P\to c_0^2 p^2 P/A_2^2$ with 
$\D q\to c_0q\D q/A_1$, $\D p \to c_0p \D p/A_2$ amount to set $f_1=h_1=1$. 
Thus, we have 
\begin{align}
\label{5DAgauge}
f_1=h_1=1\,, \qquad A \equiv \frac{[S_1(q)+S_2(p)]^{3/2}}{W_0^{1/2}(f_2-h_2)}=c_0^2  pq \,. 
\end{align}
In this gauge, the equation $R_{01}=0$ gives 
\begin{align}
\label{}
\frac 1A \partial_q (A \partial_q W_3)=\frac 1 A \partial_p (A \partial_p W_3) \,.
\end{align}
This equation is separable and yields 
\begin{align}
\label{5Dfh2}
f_2(q)=b_0 q^2+b_1 +b_2 \ln q \,, \qquad h_2(p)=-b_0 p^2+b_1' +b_2' \ln p \,,
\end{align}
where $b_0$, $b_1$, $b_2$, $b_1'$ and $b_2'$ are constant.

Next, the derivative of (\ref{5DAgauge}) yields
\begin{align}
\label{5DS1p}
S_1'(q)=\frac{2[S_1(q)+S_2(p)]}{3q} 
+\frac{2[S_1(q)+S_2(p)]f_2'(q)}{3[f_2(q)-h_2(p)]}+\frac{[S_1(q)+S_2(p)]f_0'(q)}{3[f_0(q)-h_0(p)]}\,, \\
\label{5DS2p}
S_2'(p)=\frac{2[S_1(q)+S_2(p)]}{3p} 
-\frac{2[S_1(q)+S_2(p)]h_2'(p)}{3[f_2(q)-h_2(p)]}-\frac{[S_1(q)+S_2(p)]h_0'(p)}{3[f_0(q)-h_0(p)]}\,.
\end{align}
The compatibility $\partial_p S'_1(q)=\partial_q S'_2(p)=0$ gives
\begin{align}
\label{5Dfh4eq}
\frac 13\left(\partial_p \ln \hat W_0\right)\left(\partial_q \ln \hat W_0\right)
= \frac 1p \partial_q \ln W_3+\frac 1q\partial_p \ln W_3+\frac 1{pq}\,,
\end{align}
where we have defined $\hat W_0\equiv W_0/(pqW_3)$. 
Equations $R_{00}+R_{33}=0$ and $R_{11}-R_{44}=0$ give 
\begin{align}
\label{5Df4eq}
\frac 13 \left(\partial_q \ln \hat W_0\right)^2 =&\,-\left(
2\partial_q^2 \ln W_3+(\partial_q \ln W_3)^2-(\partial_p \ln W_3)^2 -\frac 3{q^2} 
\right) \,, \\
\label{5Dh4eq}
\frac 13 \left(\partial_p \ln \hat  W_0\right)^2 =&\,-\left(
2\partial_p^2 \ln W_3-(\partial_q \ln W_3)^2+(\partial_p \ln W_3)^2 -\frac 3{p^2} 
\right) \,,
\end{align}
where we have eliminated 1st and 2nd derivatives of $S_1$ and $S_2$ by 
(\ref{5DS1p}) and (\ref{5DS2p}), and their derivatives. 
We can thus remove $\hat W_0$ by equating the product of (\ref{5Df4eq}) and  (\ref{5Dh4eq})
with the square of (\ref{5Dfh4eq}), yielding an equation involving $f_2$ and $h_2$. 
By inserting (\ref{5Dfh2}), one concludes $b_2=b_2'=0 $ and $b_1'=b_1$. 
Using the reparametrization freedom (\ref{5Daij}) of $\mathsf a_{11}$, $\mathsf a_{12}$, $\mathsf a_{22}$, 
one can set $b_1=0$ and $b_0=1$, i.e., $f_2(q)=q^2$ and $h_2(p)=-p^2$. 

With these conditions, solutions to (\ref{5Dfh4eq}),  (\ref{5Df4eq}) and  (\ref{5Dh4eq})
are found to be 
\begin{align}
\label{5Dfh4}
f_0(q)=d_1 +\frac{d_2}{q^2} \,, \qquad 
h_0(p)=d_1-\frac{d_2}{p^2} \,,
\end{align}
where $d_1$ and $d_2$ are constants. 
One can set $d_1=0$ by (\ref{shift0}). 
Substituting these into (\ref{5DAgauge}) and using the freedom (\ref{shiftS}), we find 
$d_2=1/c_0^4$ and 
\begin{align}
\label{}
S_1(q)=q^2 \,, \qquad 
S_2(p)=p^2 \,. 
\end{align}

From equation $R_{00}-R_{11}-R_{22}=-2\Lambda $, we obtain
\begin{align}
\label{}
\frac 1 A \partial_q \left[ A^{-1}\partial_q (A^2 Q)\right]+4\Lambda S_1(q)+\frac 1 A \partial_p 
\left[ A^{-1}\partial_p (A^2 P)\right]
+4\Lambda S_2 (p)=0 \,. 
\end{align}
This equation is also separable, giving 
\begin{align}
\label{}
Q''(q)+ \frac 3 q Q'(q)+4\Lambda q^2 -8c_1 =0 \,, \qquad 
P''(p)+ \frac 3 p P'(p)+4\Lambda p^2+8c_1=0\,. 
\end{align}
The general solution to these equations are
\begin{align}
\label{}
Q(q)=\frac{c_2}{q^2}+c_3+c_1 q^2-\frac 16 \Lambda q^4 \,, \qquad 
P(p)=\frac{c_4}{p^2}+c_5-c_1 p^2 -\frac 16\Lambda p^4 \,. 
\end{align}
Here $c_1$, $c_2$, $c_3$, $c_4$, $c_5$  are constants.

Finally, we shall determine $f_3(q)$ and $h_3(p)$. 
Defining
\begin{align}
\label{5DB12}
B_1 \equiv A\frac{W_3^2}{W_0}\partial_q \left(\frac{W_2}{W_3}\right)\,, \qquad 
B_2 \equiv  A\frac{W_3^2}{W_0}\partial_p \left(\frac{W_2}{W_3}\right)\,, 
\end{align}
equations $R_{02}=R_{12}=0$ lead to 
\begin{align}
\label{5DR0212}
\partial_q B_1-\frac{\partial_p W_3}{W_3} B_2 =0 \,, \qquad 
\partial_p B_2 -\frac{\partial_q  W_3}{W_3} B_1 =0 \,. 
\end{align}
Taking the derivatives of these equations, one finds
\begin{align}
\label{}
\partial _p \left(\frac{\partial_q B_1}{\partial_p W_3}\right)=-\partial_q \left(\frac{B_1}{W_3}\right)\,, \qquad 
\partial _q \left(\frac{\partial_p B_2}{\partial_q W_3}\right)=-\partial_p \left(\frac{B_2}{W_3}\right)\,.
\end{align}
It follows that there exist functions $\ma B_1(p,q)$ and $\ma B_2 (p,q)$ such that 
\begin{align}
\label{}
\partial_q \ma B_1 =\frac{\partial_q B_1}{\partial_p W_3}\,, \qquad 
\partial_p \ma B_1=-\frac{B_1}{W_3}\,, \qquad 
\partial_p \ma B_2= \frac{\partial_p B_2}{\partial_q W_3}\,, \qquad 
\partial_q \ma B_2=-\frac{B_2}{W_3}\,. 
\end{align}
Integrating the first and the third equations, 
we have 
$B_1=\partial_p W_3 \ma B_1-B_{11}'(p)$ and 
$ B_2 =\partial_q W_3 \ma B_2 -B_{22}'(q)$, where 
$B_{11}(p)$ and $B_{22}(q)$ are arbitrary functions with respect to each argument. 
Combining these with the second and the forth equations above, 
we end up with 
\begin{align}
\label{}
\ma B_1=\frac{B_{11}(p)+B_{12}(q)}{W_3} \,, \qquad 
\ma B_2=\frac{B_{22}(q)+B_{21}(p)}{W_3} \,, 
\end{align}
where $B_{12}(q)$ and $B_{21}(p)$ are arbitrary functions arising from integration.
Inserting $B_1=-W_3 \partial_p [(B_{11}(p)+B_{12}(q))/W_3]$ and 
$B_2=-W_3 \partial_q [(B_{22}(q)+B_{21}(p))/W_3]$ back into (\ref{5DR0212}), 
we obtain decoupled equations, whose integration gives
\begin{align}
\label{}
B_{12}(q)=-B_{22}(q)+b^{(0)}_1 f_2(q)+b^{(0)}_2 \,, \qquad 
B_{21}(p) =-B_{11}(p)-b^{(0)}_1 h_2(p)-b^{(0)}_2 \,, 
\end{align}
where $b^{(0)}_1$ and $b^{(0)}_2$ are constants. 
Writing $\hat B_{11}(p)=B_{11}(p)+b^{(0)}_1 h_2(p)+b^{(0)}_2$, 
equation (\ref{5DB12}) yields
\begin{align}
\label{}
\partial_q \left(\frac{B_{22}(q)-\hat B_{11}(p)}{W_3}\right)
=-\frac{A W_3}{W_0}\partial_p \left(\frac{W_2}{W_3}\right)\,, \qquad 
\partial_p \left(\frac{B_{22}(q)-\hat B_{11}(p)}{W_3}\right)
=\frac{A W_3}{W_0}\partial_q \left(\frac{W_2}{W_3}\right)\,.
\end{align}
This corresponds to the equations governing $\hat B_{11}'(p)$ and $B_{22}'(q)$.
The integrability conditions $\partial_q \hat B_{11}'(p)=0$ and $\partial_p B_{22}'(q)=0$
give rise to equations for $f_3''(q)$ and $h_3''(p)$ as
\begin{subequations}
\label{5Dfh3pp}
\begin{align}
\label{}
f_3''(q)=&\, -\frac{qW_3}{W_0}\partial_q \left(\frac{W_0}q\right)\partial_q \left(\frac{W_2}{W_3}\right)-\frac{W_2}{W_3}\partial_q^2 W_3-\partial_p W_3 \partial_p \left(\frac{W_2}{W_3}\right)\,, \\
h_3''(p)=&\, \frac{pW_3}{W_0}\partial_p \left(\frac{W_0}p\right)\partial_p \left(\frac{W_2}{W_3}\right)+\frac{W_2}{W_3}\partial_p^2 W_3+\partial_q W_3 \partial_q \left(\frac{W_2}{W_3}\right)\,.
\end{align}
\end{subequations}
Further integrability conditions
$\partial_p f_3''(q)=\partial_q h_3''(p)=0$ yield the linear system
\begin{align}
\label{}
M \vec X=0 \,, \qquad 
\label{}
M \equiv \left(
\begin{array}{cc}
M_0+M_1      &  M_2   \\
M_2      &   M_0-M_1
\end{array}
\right)\,, 
\qquad 
\vec X \equiv \left(\begin{array}{c}
\partial_p (W_2/W_3)    \\
\partial_q (W_2/W_3)  
\end{array}\right)\,, 
\end{align}
where 
\begin{align}
\label{Mi}
M_0=\left(\partial_q^2+\partial_p^2\right) \ln W_3 \,, \qquad 
M_1=\partial_p \ln W_3 \partial_p \ln \hat W_0-\partial_q \ln W_3 \partial_q \ln \hat W_0\,, 
\qquad 
M_2=\partial_ p \partial_q \ln \hat W_0\,.
\end{align}
For the existence of nontrivial solutions $\vec X\ne 0$, 
we need ${\rm det}M=0$, giving 
\begin{align}
\label{}
M_1^2+M_2^2=M_0^2 \,. 
\end{align} 
Insertion of (\ref{5Dfh2}) with $b_0=1$, $b_1=b_1'=b_2=b_2'=0$ and (\ref{5Dfh4})
into (\ref{Mi})
implies $M_0=M_1=M_2=0$, 
which assures the existence of solutions for $f_3(q)$ and $h_3(p)$. 
The governing equation (\ref{5Dfh3pp}) then simplifies to 
\begin{align}
\label{}
\frac{q^{-1}\partial_q [q^2 f_3(q)]-p^{-1}\partial_p [p^2 h_3(p)]}{p^2+q^2}={\rm const.} \,,
\end{align}
which in turn gives 
\begin{align}
\label{}
f_3(q)=e_1+e_2 q^2+\frac{e_3}{q^2}\,, \qquad 
h_3(p)=e_1-e_2p^2+\frac{e_3'}{p^2} \,,
\end{align}
where $e_1$, $e_2$, $e_3$, $e_3'$ are constants. 
Inserting all of these above into Einstein's equations, 
one arrives at $e_3'=-e_3$ and  $c_2=-c_4=c_0^4 e_3^2$. 
Finally, by using $\mathsf a_{31}$, $\mathsf a_{21}$, $\mathsf a_{33}$ in (\ref{5Daij}), 
one can set $e_3=1$, $e_1=e_2=0$, recovering the  Chen-L\"u-Pope class (\ref{CLP})
with $c_0^4=a_0$. Relabelling parameters, the structure functions take the form
\begin{align}
\label{CLP_QP}
Q(q)=a_0 q^{-2}+a_1 +a_2 q^2-\frac 16 \Lambda q^4\,, \qquad 
P(p)=-a_0 p^{-2}+a_3 -a_2 p^2-\frac 16 \Lambda  p^4 \,.
\end{align}

Above calculations proves that the only degenerate BF metric satisfying the 
vacuum Einstein's equations with a cosmological constant is uniquely determined to 
be the Chen-L\"u-Pope solution (\ref{CLP}). This includes the Myers-Perry-AdS metric. 
The coordinate transformation to the Boyer-Lindquist coordinates is 
given by 
$q=r$, $p=\sqrt{a^2\cos^2\theta+b^2\sin^2\theta}$ and
\begin{align}
\label{}
\D \tau =\D t+\frac{a^3 \Xi_a^{-1}\D \phi_1-b^3 \Xi_b^{-1}\D \phi_2}{b^2-a^2}
\,, \qquad 
\D \sigma=\frac{a\Xi_a^{-1} \D \phi_1-b\Xi_b^{-1}\D \phi_2}{b^2-a^2}\,, \qquad 
\D \chi=\frac{a\Xi_b^{-1}\D \phi_2-b\Xi_a^{-1}\D \phi_1}{ab(b^2-a^2)}\,,
\end{align}
where 
$\Xi_a=1-a^2g^2$, $\Xi_b=1-b^2g^2$. 
By defining 
\begin{align}
 \Delta_r =\frac{1}{r^2}(r^2+a^2)(r^2+b^2)(1+g^2 r^2) -2 m \,, \qquad 
\Delta_\theta =1-a^2g^2 \cos^2\theta -b^2g^2 \sin^2\theta \,,
\end{align}
the solution is brought into \cite{Hawking:1998kw}
\begin{align}
 \D s^2 =& -\frac{\Delta_r}{\rho^2 } \left(\D t -\frac {a\sin^2\theta}{\Xi_a}\D \phi_1-\frac{b\cos^2\theta }{\Xi_b}\D \phi_2 \right)^2
+\frac{\Delta_\theta \sin^2\theta }{\rho^2 } \left(-a \D t +\frac{r^2+a^2}{\Xi_a}\D \phi_1\right)^2
\nonumber  \\
&+\frac{\Delta_\theta \cos^2\theta }{\rho^2 } \left(-b \D t +\frac{r^2+b^2}{\Xi_b}\D \phi_2\right)^2 
+\rho^2 \left(\frac{\D r^2}{\Delta_r}+\frac{\D \theta ^2}{\Delta_\theta }\right) \nonumber \\ 
& +\frac{(1+g^2r^2)}{r^2\rho^2 } \left[a b \D t -\frac{b (r^2+a^2)\sin^2\theta }{\Xi_a}\D \phi_1
-\frac{a (r^2+b^2)\cos^2\theta }{\Xi_b}\D \phi_2\right]^2\,.
\end{align}
Here we have chosen
\begin{align}
\label{}
a_0=a^2b^2\,, \qquad 
a_1=a_3-2m \,, \qquad 
a_2=1+g^2(a^2+b^2)\,, \qquad 
a_3=a^2+b^2+a^2b^2g^2 \,.
\end{align}

\section{Newman-Penrose formalism}
\label{sec:NP}

In this appendix, we discuss the relationship of the Newman-Penrose quantities~\cite{Newman:1961qr}
between the seed and the background geometry. 
We follow the convention of \cite{Stephani:2003tm} with ($-1,1,1,1$) signature.

Let $\{l_\mu, n_\mu , m_\mu, \bar m_\mu\}$ span the Newman-Penrose 
null tetrad basis $g_{\mu\nu}=-2l_{(\mu }n_{\nu)}+2 m_{(\mu} \bar m_{\nu)}$, 
where $l_\mu $ and $ n_\mu $ are real null vectors with $l_\mu n^\mu=-1$, 
while $m_\mu$ is a complex null vector with $m_\mu\bar m^\mu =1$. 
Taking $l_\mu$ as the deformation vector of the Kerr-Schild, 
the deformed metric is represented by $\ti g_{\mu\nu}=g_{\mu\nu}-2S  l_\mu l_\nu$.\footnote{
In this appendix, we deliberately use a notation different from that in the main text, in order to align with the conventions of \cite{Stephani:2003tm}.
To translate back to the notation of the main text, we set $k_\mu = l_\mu$ and $H=-S$.
All symbolic tensor computations here were performed with the aid of {\sf xAct} package for {\sf Mathematica}, which also allowed us to correct typographical errors in \cite{BG1983}.
}
 
This amounts to taking the null tetrad of $\ti g_{\mu\nu}$ as 
\begin{align}
\label{}
\ti l_\mu = l_\mu \,, \qquad 
\ti n_\mu = n_\mu +S l_\mu \,, \qquad 
\ti m_\mu =m_\mu \,, 
\end{align}
with 
\begin{align}
\label{}
\ti l^\mu = l^\mu \,, \qquad 
\ti n^\mu = n^\mu -S l^\mu \,, \qquad 
\ti m^\mu =m^\mu \,.
\end{align}
Note that this is not the local Lorentz transformation to the null tetrad for a given metric, but represents the 
actual deformation of the metric. The Ricci rotation coefficients are 
varied as
\begin{align}
\label{}
\ti \kappa=&\,\kappa \,, \qquad 
\ti \sigma =\sigma \,, \qquad 
\ti \epsilon =\epsilon \,, 
\qquad 
\ti \rho =\,\rho \,, \qquad 
\ti \tau =\tau \,, \qquad \ti \pi =\pi \,, \notag \\
\ti \mu=&\,\mu+S \rho \,, \qquad \ti\lambda =\lambda+S \bar \sigma \,, \qquad 
\ti \alpha=\alpha+\frac 12 S \bar \kappa \,, \qquad 
\ti \beta =\beta +\frac 12 S \kappa \,,\\
\ti \gamma=&\,\gamma+\frac 12 (D+2\bar \epsilon+\rho-\bar \rho)S \,, \qquad 
\ti \nu=\nu+(\bar \delta +2\alpha+2\bar \beta -\pi-\bar\tau)S+S^2\bar \kappa  \,. \notag 
\end{align}
It follows that the affine-parametrized shear-free null geodesics for $g_{\mu\nu}$ 
($\kappa=\sigma={\rm Re}(\epsilon)=0$) is mapped into 
the one for $\ti g_{\mu\nu}$. Also, if $\{l^\mu, n^\mu, m^\mu, \bar m^\mu\}$ is a 
parallelly propagated frame along the geodesic $l^\mu$, i.e., $\kappa=\pi=0$, the same is true for
$\{\ti l^\mu, \ti n^\mu, \ti m^\mu, \tilde{\bar m}^\mu\}$.

The Ricci tensor components of the deformed metric are given by 
\begin{subequations}
\begin{align}
\label{}
\ti \Phi_{00}=&\,\Phi_{00}+2S |\kappa|^2 \,, 
\\
\ti \Phi_{01}=&\, \Phi_{01}+\frac 12 \left[D(\kappa S) +\kappa(D+\epsilon+3\bar \epsilon-\rho)S+\bar \kappa \sigma S\right]\,, \\
\ti \Phi_{02}=&\,\Phi_{02}+(D+3\bar\epsilon-\epsilon-2\bar \rho)(\sigma S)+
\kappa (\delta+2\bar \alpha+2\beta -\bar \pi-\tau)S +\kappa^2 S^2 \,,
\\
\ti \Phi_{11}=&\, \Phi_{11}+\frac 14 \left(D+\epsilon+\bar \epsilon+\rho+\bar \rho\right)\left(
D+2\epsilon+2\bar \epsilon-\rho-\bar \rho
\right)S+S\left(|\rho|^2+|\sigma|^2+\frac 12 \Phi_{00}+\frac 12 S|\kappa|^2\right)
\notag \\
& +\frac 14\left(\delta-\bar \alpha+\beta -\bar \pi-3\tau\right)(\bar \kappa S)
+\frac 14\left(\bar \delta- \alpha+\bar \beta - \pi-3\bar \tau\right)(\kappa S)\,,
\\
\ti \Phi_{12}=&\, \Phi_{12}+\frac 12 (\delta-\tau+\bar\alpha-\beta+\kappa S)
(D+2\bar\epsilon+\rho-\bar\rho)S-\frac 12 \Delta (\kappa S)
+\frac 12 HD(\kappa S)+D(\beta S)
\notag \\
&+\sigma (\bar \delta+2\alpha+2\bar\beta-\bar\tau-\pi+\bar\kappa S)S
+\epsilon (\delta+2\bar\alpha+2\beta-\tau-\bar\pi+\kappa S)S 
+\frac 12 \kappa(\bar\epsilon-\epsilon-\bar\rho)S^2
\notag \\
&+\left(\gamma \kappa -\rho \tau+\beta(3\bar\epsilon-\epsilon-2\bar\rho+\rho)
+\frac 12 \kappa (\gamma-\bar \gamma-\mu)-\alpha\sigma
-\frac 12 \bar\kappa (\bar\lambda+\sigma S)\right)S
 \,, 
\\
\ti \Phi_{22}=&\, \Phi_{22}+\frac 12 (\delta+\bar \alpha+3\beta-\bar\pi-\tau)(\bar \delta
+2\alpha+2\bar\beta+\pi-\bar\tau+2\bar \kappa S)S
\notag \\
&
+\frac 12 (\bar\delta+ \alpha+3\bar\beta-\pi-\bar\tau)(\delta
+2\bar\alpha+2\beta+\bar\pi-\tau+2 \kappa S)S
\notag \\
&+\frac 12 D[(\mu+\bar \mu)S]-\frac 12\Delta [(\rho+\bar \rho)S]
+2|\kappa|^2 S^3+[\Phi_{00}+\kappa(\pi-\bar \tau)+\bar\kappa(\bar\pi-\tau)]S^2
\notag \\
&+S\left[4\Lambda+\Psi_2+\bar \Psi_2+\frac 12 \big(4|\pi|^2-3(\epsilon+\bar\epsilon)(\mu+\bar \mu)
-2(\mu-\bar \mu)(\rho-\bar\rho)-(\gamma+\bar\gamma)(\rho+\bar\rho)
\big)\right]\,,\\
\ti \Lambda=&\, \Lambda-\frac 1{12}(D+\epsilon+\bar\epsilon)(D+2\epsilon+2\bar\epsilon 
-2\rho-2\bar\rho)S-\frac 16(\rho^2+\rho\bar \rho+\bar\rho^2+|\sigma|^2)S
\notag \\
&+\frac 1{12} \kappa (\bar \delta +2\alpha+2\bar \beta+\bar \kappa S)S
+\frac 1{12} \bar\kappa (\delta +2\bar\alpha+2 \beta+ \kappa S)S\,.
\end{align}
\end{subequations}
Here and only in this section, $\Lambda$ represents $\Lambda=R/(24)$ rather than the cosmological constant. 
The relation between the Weyl scalars is given by
\begin{subequations}
\begin{align}
\label{}
\ti \Psi_0=&\,\Psi_0+2\kappa^2 S \,, \\
\ti \Psi_1=&\, \Psi_1+\frac 12 D(\kappa S)+\frac 12 \kappa (D+\epsilon+3\bar \epsilon+3\rho-2\bar \rho)S -\frac 12 \bar \kappa \sigma S \,, \\
\ti \Psi_2=&\, \Psi_2+\frac 16 (D+\epsilon+\bar \epsilon+\rho-\bar \rho)(D+2\epsilon+2\bar \epsilon+3\rho-\bar\rho)S-\frac 16 \delta(\bar \kappa S)+\frac 16 \bar \delta (\kappa S)
\notag \\
&+\frac 23\kappa (\bar \delta+2\alpha+2\bar \beta-\pi-\bar \tau+\bar \kappa S )S-\frac 23 |\sigma|^2 S\notag \\
&+\frac 16 \bar \kappa (\bar \alpha-\beta-\bar \pi-\tau)S
+\frac 16 \kappa(-\alpha+\bar \beta-\pi-\bar\tau)S \,,\\
\ti \Psi_3=&\, \Psi_3+\frac 12 (\bar \delta +\alpha+\bar \beta-\bar \tau+\bar \kappa S)(D+4\epsilon
+2\bar\epsilon+\rho-\bar \rho)S -\frac 12 \Delta (\kappa S)+SD\left(\alpha+\frac 12\bar\kappa S \right)
\notag \\
& +(\rho-\epsilon)\left(\bar\delta+\alpha+2\bar\beta-\pi-\bar\tau +\frac 12\bar\kappa S\right)S
-(\tau+\beta)\bar\sigma S-\frac 12 \kappa S(\lambda +\bar\sigma S)
\notag \\
&+\frac 12 \bar \kappa (\gamma+\bar\gamma-\bar\mu-\bar\epsilon S)S
-S\alpha \bar \epsilon -2 S(\bar \delta-\alpha-\bar \beta+\pi)\epsilon\,,
\\
\ti \Psi_4=&\, \Psi_4+(\bar \delta+3\alpha+\bar \beta+\pi-\bar \tau+2\bar \kappa S)
(\bar \delta +2\alpha+2\bar \beta-\pi-\bar\tau+\bar \kappa S)S 
-\Delta (\bar \sigma S)+D(\lambda S)-2\lambda DS
\notag \\
&-\big(\bar\sigma(\mu+\bar\mu+3\gamma-\bar\gamma)+(\lambda +\bar\sigma S)(3\bar\epsilon-\epsilon+3\rho-\bar\rho)-2\nu \bar\kappa\big)S +(D \bar \sigma) S^2\,.
\end{align}
\end{subequations}

Although these transformation rules for the Newman-Penrose quantities are quite suggestive and provide useful intuition, they do not by themselves yield information about the integrability of geodesics. Such integrability essentially depends on the explicit form of the metric components, which cannot be fully captured solely by the spin coefficients and curvatures. For this reason, the discussion carried out in terms of coordinate components in the main body of the text proves to be more informative and practically useful for the present aim.

\end{document}